\documentclass[12pt, onecolumn]{IEEEtran}
\usepackage{amsmath,amsfonts}

\usepackage{siunitx}

\newcommand{\ve}{\mathbf}
\newcommand{\m}{\mathbf}

\usepackage{array}
\usepackage[caption=false,font=normalsize]{subfig}
\usepackage{textcomp}
\usepackage{stfloats}
\usepackage{url}
\usepackage{verbatim}

\usepackage{graphicx}
\usepackage{color}
\graphicspath{{./fig/}}

\usepackage[dvipsnames]{xcolor}	
\usepackage{tikz}
\usetikzlibrary{calc}
\usetikzlibrary{positioning}
\usetikzlibrary{spy}
\tikzset{>=latex}
\usepackage{pgfplots}
\pgfplotsset{compat=newest}
\pgfplotsset{plot coordinates/math parser=false}

\definecolor{myred}{RGB}{161,23,23}
\definecolor{myblue}{RGB}{23,32,161}
\definecolor{mygreen}{RGB}{66, 172,21}
\definecolor{myorange}{RGB}{245,146,33}
\definecolor{mykaki}{RGB}{209,215,38}
\definecolor{myviolet}{RGB}{205,99,243}

\usepackage{algorithm, algpseudocode}
\algrenewcommand\algorithmiccomment[1]{{\color{gray}  \hfill \small $\triangleright$ #1}}

\usepackage{tabularx}
\usepackage{booktabs}
\usepackage{makecell}
\usepackage{multirow}
\usepackage{vcell}

\usepackage{cite}
\begin{document}

\title{Neural Network Approaches for Data Estimation in Unique Word OFDM Systems}

\author{Stefan~Baumgartner~\IEEEmembership{Graduate Student Member,~IEEE}, Gerg\H{o} Bogn\'{a}r, 
Oliver Lang~\IEEEmembership{Member,~IEEE}, and Mario~Huemer~\IEEEmembership{Senior Member,~IEEE}%
\thanks{This work has been supported by the ``University SAL Labs'' initiative of Silicon Austria Labs (SAL) and its Austrian partner universities for applied fundamental research for electronic based systems, and by the Ministry of Innovation and Technology of Hungary from the National Research, Development and Innovation Fund through the TKP2021-NVA Funding Scheme under Project TKP2021-NVA-29. This paper was presented in part at the 55th Asilomar Conference on Signals, Systems, and Computers, Pacific Grove, USA, 2021~\cite{Baumgartner21_C1}.}
\thanks{Stefan Baumgartner and Mario Huemer are with the JKU LIT SAL eSPML Lab, Johannes Kepler University Linz, Austria, and with the Institute of Signal Processing, Johannes Kepler University Linz, Austria (e-mails: \{stefan.baumgartner, mario.huemer\}@jku.at).}%
\thanks{Gerg\H{o} Bogn\'{a}r is with the Department of Numerical Analysis, ELTE E\"otv\"os Lor\'{a}nd University, Budapest, Hungary (email: bognargergo@staff.elte.hu).}%
\thanks{Oliver Lang is with the Institute of Signal Processing, Johannes Kepler University Linz, Austria (email: oliver.lang@jku.at).}%
}



\maketitle

\begin{abstract}
Data estimation is conducted with model-based estimation methods since the beginning of digital communications. 
However, motivated by the growing success of machine learning, current research focuses on replacing model-based data estimation methods by data-driven approaches, mainly neural networks (NNs). In this work, we particularly investigate the incorporation of existing model knowledge into data-driven approaches, which is expected to lead to complexity reduction and / or performance enhancement. We describe three different options, namely ``model-inspired'' pre-processing, choosing an NN architecture motivated by the properties of the underlying communication system, and inferring the layer structure of an NN with the help of model knowledge. Most of the current publications on NN-based data estimation deal with general multiple-input multiple-output communication (MIMO) systems. In this work, we investigate NN-based data estimation for so-called unique word orthogonal frequency division multiplexing (UW-OFDM) systems. We highlight differences between UW-OFDM systems and general MIMO systems one has to be aware of when using NNs for data estimation, and we introduce measures for successful utilization of NN-based data estimators in UW-OFDM systems. Further, we investigate the use of NNs for data estimation when channel coded data transmission is conducted, and we present adaptions to be made, such that NN-based data estimators provide satisfying performance for this case. We compare the presented NNs concerning achieved bit error ratio performance and computational complexity, we show the peculiar distributions of their data estimates, and we also point out their downsides compared to model-based equalizers.
\end{abstract}


\section{Introduction}
On the receiver side of wireless digital communication systems, data estimation, also referred to as equalization, is conducted to reconstruct the transmitted data that have been disturbed during transmission. Traditionally, this task is accomplished with model-based estimation methods. That is, the data transmission is described by physical and mathematical models, such that on basis of these models statistical estimation methods can be developed to estimate the transmitted data. This established approach has many advantages, e.g., the derived estimation methods are well-interpretable, and often performance bounds can be derived. However, there are also some downsides. Model-based estimation methods yielding optimal performance are generally computationally infeasible, which requires resorting to less complex, suboptimal methods in practice. Furthermore, modeling inaccuracies may lead to severe performance degradation, and the empirical statistical behavior of available data cannot be utilized for improving the estimation results. With data-driven machine learning methods, 
some of the aforementioned issues of model-based approaches can be resolved. Hence, employing data-driven methods, particularly neural networks (NNs), for equalization is a focus of current research~\cite{OShea17, Samuel19, He20, He18, Liao20, Khani20, Shlezinger20, Shlezinger21, Pratik21, Luong22}. The employed NNs, which are known to be universal function approximators~\cite{Hornik91}, should approximate the optimal data estimator function. Ideally, the developed NN-based data estimators exhibit a low computational complexity and require a low amount of training data. However, this is a major challenge, since most of the currently known standard NNs usually have a large number of trainable parameters, leading to a large amount of required training data and a high inference complexity. One approach for tackling this problem is to design the layer architecture of an NN based on the model of the data transmission. In the context of such model-inspired NN layer structures, the concept of deep unfolding~\cite{Hershey14} is worth mentioning. There, the iterations of an iterative model-based algorithm are unfolded to layers of an NN, where free parameters or even whole parts of the model-based inference structure are replaced by trainable parameters or modules, respectively, that are optimized with the help of training data. NN-based data estimators which are deduced by deep unfolding are, e.g., DetNet~\cite{Samuel19}, OAMP-Net~\cite{Liao20}, MMNet~\cite{Khani20}, ViterbiNet~\cite{Shlezinger20}, or DeepSIC~\cite{Shlezinger21}, to name just a few. Besides deep unfolding, there are also other approaches for model knowledge aided data-driven methods for data estimation. In~\cite{Mohammadkarimi19}, a deep learning aided sphere decoder is proposed, where the radius of the decoding hypersphere is learned. Further, in~\cite{Pratik21}, RE-MIMO, an NN-based equalizer for multiple-input multiple-output (MIMO) systems with a variable number of transmitters is published, consisting of three modules, where the NN architecture of each module is chosen motivated by model-based considerations. 

An important distinction between NN-based equalizers can be made by considering their generalization ability regarding different channels. While the aforementioned DetNet and OAMP-Net are trained with an ensemble of different channels, representing samples from a statistical channel model, and use the actual channel realization as an input, MMNet, ViterbiNet, and DeepSIC are trained for a single channel realization. The former NNs can be trained completely offline\footnote{We assume that all channel realizations during inference exhibit the same statistics as those used for training. In case the channel statistics change, the NNs have to be retrained.}, but generally feature a higher inference complexity than the latter. The latter have to be (re-)trained in an online manner, however, they usually contain fewer trainable parameters, leading to a faster training convergence and a smaller amount of required training data. In this paper, we focus on NNs that are trained offline with a multitude of different channels and use the current channel realization as an input during inference. 
 
In this work, we illustrate different approaches for obtaining model-inspired NN-based data estimation methods, namely, by model-aware pre-processing, and by employing NN architectures tailored for the structure of the model describing the data transmission. More specifically, we consider NN-based equalization for communication systems employing the so-called unique word orthogonal frequency division multiplexing (UW-OFDM) signaling scheme~\cite{HuemHofbHub10}. UW-OFDM is an alternative to the popular cyclic prefix (CP)-OFDM, which allows, among others, achieving a better bit error ratio (BER) performance than with CP-OFDM, however, at the cost of a higher equalization complexity. That is, while a low-complex single-tap (per subcarrier) equalizer provides already optimal performance for CP-OFDM, this is not the case for UW-OFDM. More specifically, for UW-OFDM more complex linear equalizers are required, and the performance can be improved even further by nonlinear data estimators. This motivates the investigation of NNs as an alternative to model-based equalizers in UW-OFDM systems. This work gives an extensive investigation of NN approaches for data estimation in UW-OFDM systems, however, many of the ideas are quite general, and can also be transferred to other digital communication waveforms. We naturally highlight the advantages of the presented ideas, but intentionally also discuss the downsides and open challenges that still need to be overcome in NN-based data estimation and / or model-inspired NN-based equalizers for future digital communication systems.

We start by applying DetNet, which has originally been proposed for a MIMO system, to the considered UW-OFDM systems, and point out the required adaptions. Further, we employ a fully-connected NN (FCNN) as a data estimator, whereby a model-aware data pre-processing scheme is applied to its input data. As a third NN-based equalizer, we propose an NN based on the transformer architecture~\cite{Vaswani17}, which utilizes self-attention to exploit correlations in the data for improving the equalization performance. The majority of available publications on NN-based data estimators assume a MIMO system model -- often with data transmission over an uncorrelated Rayleigh fading channel. Due to the different system properties of UW-OFDM systems in comparison to MIMO systems over uncorrelated Rayleigh fading channels, pre-processing steps are required to obtain well-performing NN-based equalizers, which we detail in this paper. We compare the NN-based approaches with model-based methods in terms of performance and complexity. We conduct our investigations for both channel coded and uncoded data transmission, where the former case has rarely been covered in publications on NN-based data estimators yet. For channel coded transmission, the equalizers have to provide reliability information about their estimates. It turns out, that NN-based data estimators tend to be overconfident in their decisions, which impairs the overall system performance. We suggest a measure that can be conducted to counteract the overconfidence of the NNs, which allows achieving approximately the same BER performance as with an optimal equalizer. Furthermore, we plot the empirical distributions of the estimates of model-based and NN-based equalizers in the in-phase/quadrature-phase (I/Q)-diagram, which highlights peculiarities of some of the considered equalizers. 

The remainder of this paper is structured as follows: we start by reviewing the UW-OFDM signaling scheme in Sec.~\ref{sec:Preliminaries}. In Sec.~\ref{sec:Model-Based_Data_Estimation}, we present optimal and suboptimal model-based data estimation methods, and we visualize their decision boundaries in a toy example. We address the NN-based equalizers, as well as the utilized data normalization scheme, in Sec.~\ref{sec:Neural_Network_Based_Data_Estimation}. In Sec.~\ref{sec:Results}, we provide BER performance results for both channel coded and uncoded data transmission, we conduct a complexity analysis, and we compare the distributions of the estimates provided by model-based and NN-based equalizers. Finally, we conclude with our findings in Sec.~\ref{sec:Conclusion}.

\subsection*{Notation}
Throughout this paper, the $i$th element of a vector $\ve{x}$, the element in the $i$th row and the $j$th column of a matrix $\m{X}$, and the $i$th row of a matrix $\m{X}$ are denoted as $x_i$, $[\m{X}]_{ij}$, and $[\m{X}]_{i,*}$, respectively. The operators $\text{Re}\{.\}$ and $\text{Im}\{.\}$ deliver the real and the imaginary part of a complex-valued quantity, $(.)^T$ and $(.)^H$ indicate the transposition and the conjugate transposition of a vector/matrix, respectively. Furthermore, $p(.)$, $p[.]$, $\text{Pr}(.)$, $p[a|b]$, $p[a=\tilde{a}]$, and $E_a[.]$ describe the probability density function (PDF) of a continuous random variable, the probability mass function (PMF) of a discrete random variable, the probability operator, a conditional PMF of the random variable $a$ given $b$, a PMF evaluated at the value $\tilde{a}$, and the expectation operator averaging over the PDF/PMF of $a$, respectively. The subscript of the expectation operator is omitted, when the averaging PDF/PMF is clear from context.

\section{Preliminaries}
\label{sec:Preliminaries}
In this section, we describe the basics of UW-OFDM. For more detailed information on UW-OFDM, we refer to~\cite{HuemHofbHub10, Huemer11, HuemHofbHub12, HuemHofbOnicHub14}. The UW-OFDM signaling scheme mainly exhibits two differences from CP-OFDM. Firstly, a deterministic sequence, the so-called UW, is employed as a guard interval. Secondly, the guard interval is part of a UW-OFDM time domain symbol resulting from an inverse discrete Fourier transform (IDFT) operation. That is, a guard interval is not removed on receiver side, but is transformed to frequency domain together with the preceding payload. With this approach, redundancy in frequency domain is introduced, which can be exploited beneficially for spectral shaping~\cite{Rajabzadeh14}, and for achieving a better BER performance~\cite{HuemHofbHub10} than with CP-OFDM, however, at the cost of receiver complexity. In the following, we elucidate the data transmission in a UW-OFDM system and its associated system model.

As in CP-OFDM, the data symbols, drawn from a phase-shift keying (PSK) or quadrature amplitude modulation (QAM) alphabet\footnote{In this paper, the alphabet is assumed to be QPSK (quadrature PSK).} $\mathbb{S}^\prime$, are defined in frequency domain. In contrast to a CP-OFDM symbol, a UW-OFDM symbol $\tilde{\ve{x}}\in\mathbb{C}^{N}$, containing $N_{\text{d}}$ data symbols $d^\prime\in\mathbb{S}^\prime$, has to fulfill some conditions. To reveal the conditions on a UW-OFDM symbol, we consider the structure and the generation of a UW-OFDM time domain symbol $\ve{x}_{\text{t}}\in\mathbb{C}^{N}$ of length $N$. In a first step, a time domain symbol $\ve{x}$ is generated that consists of payload data $\ve{x}_\text{pl}$, and a succeeding sequence of zeros with length $N_{\text{u}}$, i.e., $\ve{x} = [\ve{x}_\text{pl}^T \quad \ve{0}^T]^T$. The requested structure of $\ve{x}$ imposes the condition $\m{F}_N^{-1}\tilde{\ve{x}} = [\ve{x}_\text{pl}^T\quad \ve{0}^T]^T$ on the corresponding UW-OFDM symbol $\tilde{\ve{x}}$ in frequency domain, where $\m{F}_N^{-1}$ is the $N$-point IDFT matrix. To fulfill this constraint, the number of data symbols $N_{\text{d}}$ per UW-OFDM symbol has to be at least by $N_{\text{u}}$ smaller than the length $N$ of a UW-OFDM symbol, reduced by the number of zero subcarriers $N_{\text{z}}$, i.e., $N_{\text{d}}~\leq~N - N_{\text{z}} - N_{\text{u}}$. Throughout this paper, we consider the case\footnote{In case of additionally employing $N_{\text{p}}$ pilot subcarriers, $N_{\text{d}}$ has to be further reduced by $N_{\text{p}}$. For simplicity, we omit the inclusion of pilot subcarriers in this derivation. Details on including pilot subcarriers in UW-OFDM symbols, can be found in~\cite{Hofbauer20} and \cite{Hofbauer16}.} $N_{\text{d}} = N - N_{\text{z}} - N_{\text{u}}$. The generation of a UW-ODFM symbol is described by $\tilde{\ve{x}} = \m{B}\m{G}\ve{d}^\prime$, where $\ve{d}^\prime\in\mathbb{S}^{\prime N_{\text{d}}}$ is the data vector, $\m{B}\in\{0,1\}^{N\times (N_{\text{d}} + N_{\text{u}})}$ models the optional insertion of zero subcarriers, and $\m{G}\in\mathbb{C}^{(N_{\text{d}}+N_{\text{u}})\times N_{\text{d}}}$ is the so-called generator matrix. The generator matrix $\m{G}$ can be decomposed into $\m{G} = \m{A}\begin{bmatrix}
\m{I}\\ \m{T}
\end{bmatrix}$, with the $N_{\text{d}}\times N_{\text{d}}$ identity matrix $\m{I}$, and an appropriately chosen matrix $\m{T}\in\mathbb{C}^{N_{\text{u}}\times N_{\text{d}}}$, ensuring $N_{\text{u}}$ trailing zeros in the UW-OFDM time domain symbol. The matrix $\m{A}\in\mathbb{R}^{(N_{\text{d}}+N_{\text{u}})\times (N_{\text{d}}+N_{\text{u}})}$, in turn, can be any non-singular matrix, which can be chosen according to the so-called systematic or non-systematic UW-OFDM signaling scheme. In this work, the non-systematic approach is used, where $\m{A}$ is optimized for the BER performance of the linear minimum mean square error (LMMSE) data estimator as in~\cite{HuemHofbHub12}. In case $\m{A}$ is chosen to be a permutation matrix placing the data symbols and the redundant values on their intended subcarrier position, the signaling scheme is termed systematic UW-OFDM. For further details on systematic and non-systematic UW-OFDM, we refer to~\cite{HuemHofbHub10, Huemer11, HuemHofbHub12, HuemHofbOnicHub14}.

The last step on transmitter side is generating a transmit symbol $\ve{x}_\text{t}$ by inserting the deterministic UW $\ve{x}_\text{u}\in\mathbb{C}^{N_{\text{u}}}$ at the position of the zero sequence of the UW-OFDM time domain symbol, i.e., $\ve{x}_{\text{t}} = \ve{x} + [\ve{0}^T\quad \ve{x}_{\text{u}}^T]^T$. After transmission of $\ve{x}_{\text{t}}$ over a multipath channel and additional disturbance by additive white Gaussian noise (AWGN), the corresponding received vector is transformed to frequency domain, and the zero subcarriers are removed. The resulting downsized vector $\ve{y}_{\text{d}}$ follows to 
\begin{gather}
\ve{y}_{\text{d}} = \tilde{\m{H}}\m{G}\ve{d}^\prime + \tilde{\m{H}}\m{B}^T\tilde{\ve{x}}_\text{u} + \m{B}^T\m{F}_N\ve{n}\,,
\end{gather}
where the diagonal matrix $\tilde{\m{H}}\in\mathbb{C}^{(N_{\text{d}}+N_{\text{u}})\times (N_{\text{d}}+N_{\text{u}})}$ contains the sampled channel frequency response excluding the positions of the zero subcarriers, $\m{F}_N$ is the $N$-point discrete Fourier transform (DFT) matrix, $\tilde{\ve{x}}_{\text{u}} = \m{F}_N\,[\ve{0}^T\quad \ve{x}_{\text{u}}^T]^T$ denotes the UW in frequency domain, and $\ve{n}\sim\mathcal{C N}(\ve{0}, \sigma_{\text{n}}^2\m{I})$ is circularly symmetric complex white Gaussian noise, where $\sigma_{\text{n}}^2$ is the variance of the AWGN in time domain. 

Removing the influence of the known UW on $\ve{y}_{\text{d}}$ yields the equivalent complex baseband system model
\begin{gather}
\ve{y}^\prime = \ve{y}_{\text{d}} - \tilde{\m{H}}\m{B}^T\tilde{\ve{x}}_\text{u} = \m{H}^\prime\ve{d}^\prime + \ve{w}^\prime\,,\label{eq:system_model}
\end{gather}
with $\m{H}^\prime = \tilde{\m{H}}\m{G}$ and $\ve{w}^\prime\sim\mathcal{C N}(\ve{0}, N\sigma_{\text{n}}^2\m{I})$. 

In case of channel coded data transmission, reliability information of the estimates, also referred to as soft information or soft decision estimates, has to be provided, e.g., in form of log-likelihood ratios (LLRs)
\begin{gather}
L_{ji} = \ln\left(\frac{\text{Pr}(b_{ji}=1|\ve{y}^\prime)}{\text{Pr}(b_{ji}=0|\ve{y}^\prime)}\right)\,,\label{eq:LLR_definition}
\end{gather}
with $L_{ji}$ being the LLR of the $j$th bit $b_{ji}$ of the $i$th data symbol, $j\in\{0, ..., \log_2(|\mathbb{S}^\prime|-1)\}$, $i\in\{0, ..., N_{\text{d}}-1\}$. The LLRs serve as input for the channel decoder. For uncoded data transmission, the data symbol estimates are sliced to the nearest symbol in the symbol alphabet, which is also termed hard decision estimation.

\section{Model-Based Data Estimation}
\label{sec:Model-Based_Data_Estimation}
In this section, we review some traditional, model-based approaches for equalization. The aim is to estimate the data vector $\ve{d}^\prime$ based on the received vector $\ve{y}^\prime$, the channel state information in form of the matrix $\m{H}^\prime$, and the system model~\eqref{eq:system_model}. We start by elaborating on optimal estimators, which are, however, in general computationally infeasible. Consequently, one usually has to resort to suboptimal estimation methods in practice. We describe two state-of-the-art suboptimal estimators, where one is a linear and the other one is a non-linear estimator. Further, we present the decision boundaries of the aforementioned equalizers in a toy example to visualize their differences.

\subsection{Bit-Wise Maximum A-Posteriori Estimator}
The optimal estimator in terms of the BER performance is the bit-wise maximum a-posteriori (MAP) estimator~\cite{Stroem16}, yielding the bit value featuring the highest probability for a given received vector $\ve{y}^\prime$ according to
\begin{align}
\hat{b}_{ji} &= \underset{\tilde{b}\in\{0,1\}}{\arg\max}\,p[b_{ji}=\tilde{b}|\ve{y}^\prime] = \underset{\tilde{b}\in\{0,1\}}{\arg\max}\,\sum_{\ve{d}^{\prime\prime}\in\mathcal{S}_{ji}^{(\tilde{b})}}p(\ve{y}^\prime|\ve{d}^{\prime\prime})\nonumber\\ &= \underset{\tilde{b}\in\{0,1\}}{\arg\max}\sum_{\ve{d}^{\prime\prime}\in\mathcal{S}_{ji}^{(\tilde{b})}}\exp\Big(-\frac{1}{N\sigma_{\text{n}}^2}||\ve{y}^\prime-\m{H}^\prime\ve{d}^{\prime\prime}||_2^2\Big)\,,\label{eq:bit-wise_MAP_estimator}
\end{align}
where $\mathcal{S}_{ji}^{(\tilde{b})}\subset\mathbb{S}^{\prime N_{\text{d}}}$ denotes the set of data vectors with the bit $b_{ji}$ fixed to the value $\tilde{b}\in\{0,1\}$. For the second step in~\eqref{eq:bit-wise_MAP_estimator}, the data symbols in the data vector are assumed to be independent and identically distributed (i.i.d.) with a uniform prior probability.

\subsection{Vector Maximum Likelihood Estimator}
\label{ssec:Vector_Maximum_Likelihood_Estimator}
The estimated data vector $\hat{\ve{d}}^\prime$ produced by the vector maximum likelihood (ML) estimator maximizes the likelihood function $p(\ve{y}^\prime|\ve{d}^{\prime\prime})$, whereby all possible data vectors $\ve{d}^{\prime\prime}\in\mathbb{S}^{\prime N_{\text{d}}}$ are considered. Since we assume to have i.i.d. data symbols in the data vector, the vector ML estimator coincides with the vector MAP estimator. The vector ML estimator is given by
\begin{gather}
\hat{\ve{d}}^\prime = \underset{\ve{d}^{\prime\prime}\in\mathbb{S}^{\prime N_{\text{d}}}}{\arg\max}\,p(\ve{y}^\prime|\ve{d}^{\prime\prime}) = \underset{\ve{d}^{\prime\prime}\in\mathbb{S}^{\prime N_{\text{d}}}}{\arg\min}\,||\ve{y}^\prime - \m{H}^\prime\ve{d}^{\prime\prime}||_2^2\,.\label{eq:vec_ML}
\end{gather}

In literature, this estimator is often considered to be the optimal equalizer. In fact, it is optimal with respect to the error probability of the data vector estimate~\cite{Stroem16}, but not with respect to the BER, which is the usual figure of merit in communications. By examining~\eqref{eq:vec_ML}, a noteworthy peculiarity of the vector ML estimator can be observed, namely, this estimator does not depend on the noise variance $\sigma_{\text{n}}^2$, which is in contrast to the bit-wise MAP estimator~\eqref{eq:bit-wise_MAP_estimator}. 

\subsection{Minimum Mean Square Error Estimator}
\label{ssec:Minimum_Mean_Square_Error_Estimator}
The nonlinear MMSE estimator is, in contrast to the ML and the MAP estimators, very rarely regarded in communications literature. Especially when it comes to NN-based data estimators, which try to approximate the nonlinear MMSE estimator, we believe that a detailed consideration of the nonlinear MMSE estimator is quite meaningful. 

When employing the Bayesian mean square error $E_{\ve{y}^\prime,\ve{d}^\prime}[||\hat{\ve{d}}^\prime - \ve{d}^\prime||_2^2]$ as a performance measure, the minimum mean square error (MMSE) estimator is the optimal estimator. The MMSE estimator is obtained by computing the mean of the posterior PMF~\cite{Kay93}, i.e.,
\begin{align}
\hat{\ve{d}}^\prime &= E_{\ve{d}^\prime|\ve{y}^\prime}[\ve{d}^\prime|\ve{y}^\prime] = \sum_{\ve{d}^{\prime\prime}\in\mathbb{S}^{\prime N_{\text{d}}}}\ve{d}^{\prime\prime} p[\ve{d}^{\prime\prime}|\ve{y}^\prime]\nonumber\\ &= \frac{\sum\limits_{\ve{d}^{\prime\prime}\in\mathbb{S}^{\prime N_{\text{d}}}}\ve{d}^{\prime\prime}\,\exp(-\frac{1}{N\sigma_{\text{n}}^2}||\ve{y}^\prime-\m{H}^\prime\ve{d}^{\prime\prime}||_2^2)}{\sum\limits_{\ve{d}^{\prime\prime}\in\mathbb{S}^{\prime N_{\text{d}}}} \exp(-\frac{1}{N\sigma_{\text{n}}^2}||\ve{y}^\prime-\m{H}^\prime\ve{d}^{\prime\prime}||_2^2)}\,,\label{eq:MMSE_estimator}
\end{align}
where again a uniform prior probability distribution of the data vectors is assumed. 

Interestingly, as shown in Appendix~\ref{sec:Equivalence_MMSE_Bit-Wise_MAP_Hard_Decision_Estimates}, for a QPSK modulation alphabet (which is employed as modulation alphabet in this paper) the hard decision estimates of the MMSE estimator coincide with those of the bit-wise MAP estimator. Hence, the MMSE estimator also serves as a benchmark for the best BER performance achievable. For higher-order modulation alphabets, e.g., 16-QAM or 64-QAM, the MMSE has to be formulated for the transmitted bit vector (instead of the complex-valued data symbol vector) for obtaining optimal BER performance. 

\subsubsection*{Reliability Information for MMSE Estimates}
As obvious from \eqref{eq:LLR_definition}, the posterior probabilities $\text{Pr}(b_{ji}=1|\ve{y}^\prime)$ and $\text{Pr}(b_{ji}=0|\ve{y}^\prime)$ have to be determined to obtain the desired LLRs $L_{ji}$. 
For the employed QPSK modulation alphabet, the LLRs $L_{0i}$ and $L_{1i}$, corresponding to the zeroth and the first bit of the $i$th data symbol $d_i^\prime$, respectively, can be computed on basis of the MMSE estimates $\hat{d}_i^\prime$ with low complexity, which is presented in the following. To this end, let us consider the QPSK bit-to-symbol mapping $(b_{1i}b_{0i}) \mapsto d_i^\prime$, where the bits $b_{0i}$ and $b_{1i}$ are mapped to the real part and the imaginary part of $d_i^\prime$, respectively. The bit values $0$ and $1$ are mapped to the symbol values $-\rho$ and $\rho$, respectively, with the energy normalization factor $\rho=1/\sqrt{2}$.
Hence, as given in \eqref{eq:ith_MMSE_estimate_QPSK}, the real part of the $i$th MMSE estimate follows to
\begin{align}
\text{Re}\{\hat{d}_i^\prime\} &= E_{d_i^\prime|\ve{y}^\prime}[\text{Re}\{d_i^\prime\}|\ve{y}^\prime]\nonumber\\ &= \rho \text{Pr}(\text{Re}\{d_i^\prime\} = \rho|\ve{y}^\prime) -\rho \text{Pr}(\text{Re}\{d_i^\prime\} = -\rho|\ve{y}^\prime)\nonumber\\ &= \rho\text{Pr}(b_{0i} = 1|\ve{y}^\prime) -\rho\text{Pr}(b_{0i} = 0|\ve{y}^\prime)\,.\label{eq:ith_MMSE_Re_estimate_QPSK}
\end{align}
Since $\text{Pr}(b_{0i}~=~0|\ve{y}^\prime)+\text{Pr}(b_{0i}~=~1|\ve{y}^\prime)=1$, \eqref{eq:ith_MMSE_Re_estimate_QPSK} can be expressed as 
\begin{gather}
\text{Re}\{\hat{d}_i^\prime\} = \rho(2\text{Pr}(b_{0i} = 1|\ve{y}^\prime) - 1)\,,\label{eq:ith_MMSE_Re_estimate_QPSK_Pr1}
\end{gather}
or as 
\begin{gather}
\text{Re}\{\hat{d}_i^\prime\} = \rho(1-2\text{Pr}(b_{0i} = 0|\ve{y}^\prime))\,.\label{eq:ith_MMSE_Re_estimate_QPSK_Pr0}
\end{gather}
By rearranging \eqref{eq:ith_MMSE_Re_estimate_QPSK_Pr1} and \eqref{eq:ith_MMSE_Re_estimate_QPSK_Pr0} with respect to the posterior probabilities, and plugging the results into the LLR definition~\eqref{eq:LLR_definition} yields 
\begin{gather}
L_{0i} = \ln\bigg(\frac{\rho + \text{Re}\{\hat{d}_i^\prime\}}{\rho - \text{Re}\{\hat{d}\}}\bigg)\,,\quad \text{and} \quad L_{1i} = \ln\bigg(\frac{\rho + \text{Im}\{\hat{d}_i^\prime\}}{\rho - \text{Im}\{\hat{d}\}}\bigg)\,,
\end{gather}
where for obtaining $L_{1i}$  the same steps as above have to be conducted for the imaginary part of $\hat{d}_i^\prime$.

\subsection{Linear Minimum Mean Square Error Estimator}
\label{ssec:LMMSE}
The aforementioned optimal equalizers all suffer from a complexity that is exponential in the length of the data vector. To obtain low-complex equalizers, one can constrain the estimator to be linear. The best linear estimator in terms of the Bayesian mean square error is the LMMSE estimator. By applying the Bayesian Gauss-Markov theorem~\cite{Kay93} to \eqref{eq:system_model}, the LMMSE estimator follows to
\begin{gather}
\hat{\ve{d}}^\prime = \m{E}_{\text{LMMSE}}\ve{y}^\prime = \left(\m{H}^{\prime H}\m{H}^\prime + \frac{N\sigma_{\text{n}}^2}{\sigma_{\text{d}}^2}\m{I}\right)^{-1}\m{H}^{\prime H}\ve{y}^\prime\,,\label{eq:LMMSE_estimator}
\end{gather}
where $\sigma_{\text{d}}^2$ is the variance of the data symbols, and $\m{E}_{\text{LMMSE}}$ is the LMMSE estimator matrix. 

\subsubsection*{Reliability Information for LMMSE Estimates}
With LMMSE estimates $\hat{d}_i^\prime$ at hand, the LLRs can be computed by evaluating the alternative LLR definition
\begin{gather}
L_{ji}^{\text{LMMSE}} = \ln\left(\frac{\text{Pr}(b_{ji}=1|\hat{d}_i^\prime)}{\text{Pr}(b_{ji}=0|\hat{d}_i^\prime)}\right)\,.\label{eq:LLR_definition_LMMSE}
\end{gather}
By assuming a Gaussian conditional distribution $p(\hat{d}_i^\prime|d_i^\prime)$ for the LMMSE estimates (which is valid for large $N_{\text{d}}$ following central limit theorem arguments), it can be shown~\cite{Haselmayr15}, that~\eqref{eq:LLR_definition_LMMSE} is equivalent to the LLR definition~\eqref{eq:LLR_definition}. Following the derivation described in~\cite{Hofbauer16}, the LLRs for the zeroth and the first bit are given by
\begin{gather}
L_{0i} = \frac{4 \text{Re}\{\hat{d}_i\}\alpha_i\rho}{\sigma_i^2}\quad\text{and} \quad L_{1i} = \frac{4 \text{Im}\{\hat{d}_i\}\alpha_i\rho}{\sigma_i^2}\,,\label{eq:LLRs_LMMSE}
\end{gather}
respectively, where $\sigma_i^2 = \ve{e}_i^H(\sigma_{\text{d}}^2\bar{\m{H}}_i^\prime\bar{\m{H}}_i^{\prime H} + N\sigma_{\text{n}}^2\m{I})\ve{e}_i$, $\alpha_i~=~\ve{e}_i^H\ve{h}_i^\prime$, $\ve{e}_i^H$ is the $i$th row of $\m{E}_{\text{LMMSE}}$, $\ve{h}_i^\prime$ denotes the $i$th column of $\m{H}^\prime$, and $\bar{\m{H}}_i^\prime$ is $\m{H}^\prime$ without the $i$th column. 

\subsection{Decision-Feedback Equalizer}
A performance-complexity trade-off is provided by the decision-feedback equalizer (DFE), which is summarized in Alg.~\ref{alg:DFE}. In this iterative method, LMMSE estimation of a single data symbol is conducted in every iteration. As a decision criterion which data symbol is estimated in the $k$th iteration, we use the diagonal of the LMMSE error covariance matrix $\m{C}_{\text{ee},k}$, containing the error variances of the LMMSE estimates. That is, in a single iteration the data symbol corresponding to the smallest error variance is estimated (cf. Alg.~\ref{alg:DFE}, line~\ref{alg:DFE_Cee_min}), followed by updating the system model in form of removing the influence of the hard decision estimate $\lfloor \hat{d}_i\rceil$ from the received vector (Alg.~\ref{alg:DFE}, line~\ref{alg:DFE_remove_estim_influence}), and by deleting the appropriate column from the system matrix $\m{H}_k^\prime$ of the $k$th iteration (Alg.~\ref{alg:DFE}, line~\ref{alg:DFE_remove_column_from_H}). 

\begin{algorithm}[t]
\caption{Decision Feedback Equalization}
\label{alg:DFE}
\begin{algorithmic}[1]
\Function{DFE}{$\m{H}^\prime$, $\ve{y}^\prime$, $\sigma_{\text{n}}^2$, $\sigma_{\text{d}}^2$}
\State $\m{H}_0^\prime \gets \m{H}^\prime$, $\ve{y}_0^\prime \gets \ve{y}^\prime$
\State $\ve{r}_{\text{idx}} = \begin{bmatrix}0, 1, \dots, N_{\text{d}}-1\end{bmatrix}^T$\Comment{residual index vector}
\For{$k = 0, ..., N_{\text{d}}-1$}
\State $\m{A}_k \gets \left(\m{H}_k^{\prime H}\m{H}_k^\prime + \frac{N\sigma_{\text{n}}^2}{\sigma_{\text{d}}^2}\m{I}\right)^{-1}$\label{alg:DFE_inverse_term}
\State $\m{C}_{\text{ee},k} \gets N\sigma_{\text{n}}^2\m{A}_k$\Comment{error covariance matrix}\label{alg:DFE_Cee}
\State $j \gets \arg\min_m\,[\m{C}_{\text{ee},k}]_{mm}$\label{alg:DFE_Cee_min}
\State $\ve{e}_k^H \gets \left[\m{A}\right]_{j,*}\m{H}_k^{\prime H}$\label{alg:DFE_estimator_vector}
\State $i \gets r_{\text{idx},j}$\Comment{data symb. index to be estim.}
\State \textbf{remove} $r_{\text{idx},j}$ from $\ve{r}_{\text{idx}}$
\State $\hat{d}_i^\prime \gets \ve{e}_k^H\ve{y}_k^\prime$\Comment{data symb. estimation}
\State $\ve{h}_j \gets \left[\m{H}_k^\prime\right]_{*,j}$
\State $\m{H}_{k+1}^\prime \gets \m{H}_k^\prime\,\backslash\,\ve{h}_j$ \Comment{remove $\ve{h}_j$ from $\m{H}_k^\prime$}\label{alg:DFE_remove_column_from_H}
\State $\ve{y}_{k+1}^\prime \gets \ve{y}_k^\prime - \ve{h}_j\left\lfloor \hat{d}_i^\prime \right\rceil$\Comment{rem. influence of estim.}\label{alg:DFE_remove_estim_influence}
\EndFor
\State\Return{$\hat{\ve{d}}^\prime$}
\EndFunction
\end{algorithmic}
\end{algorithm}

\subsubsection*{Reliability Information for the DFE}
Due to the non-linear iterative equalization process, the best BER results for channel coded data transmission are obtained by incorporating channel decoding into the iterations of the DFE. 
However, in this work, we do not consider information feedback from the channel decoder to any of the regarded equalizers. As a circumvention, we utilize the LLRs of the LMMSE data symbol estimation in every iteration as reliability information of the DFE. Hence, the LLRs $L_{0i}$ and $L_{1i}$ corresponding to the data symbol estimate $\hat{d}_i^\prime$ estimated in the $k$th iteration are computed as given in~\eqref{eq:LLRs_LMMSE}, whereby $\m{H}^\prime$ is replaced by $\m{H}_k^\prime$. 

\subsection{Decision Boundaries of Model-Based Equalizers}

\begin{figure*}[b]
\centering
\subfloat[MMSE]{
\centering
\includegraphics[width=0.17\textwidth]{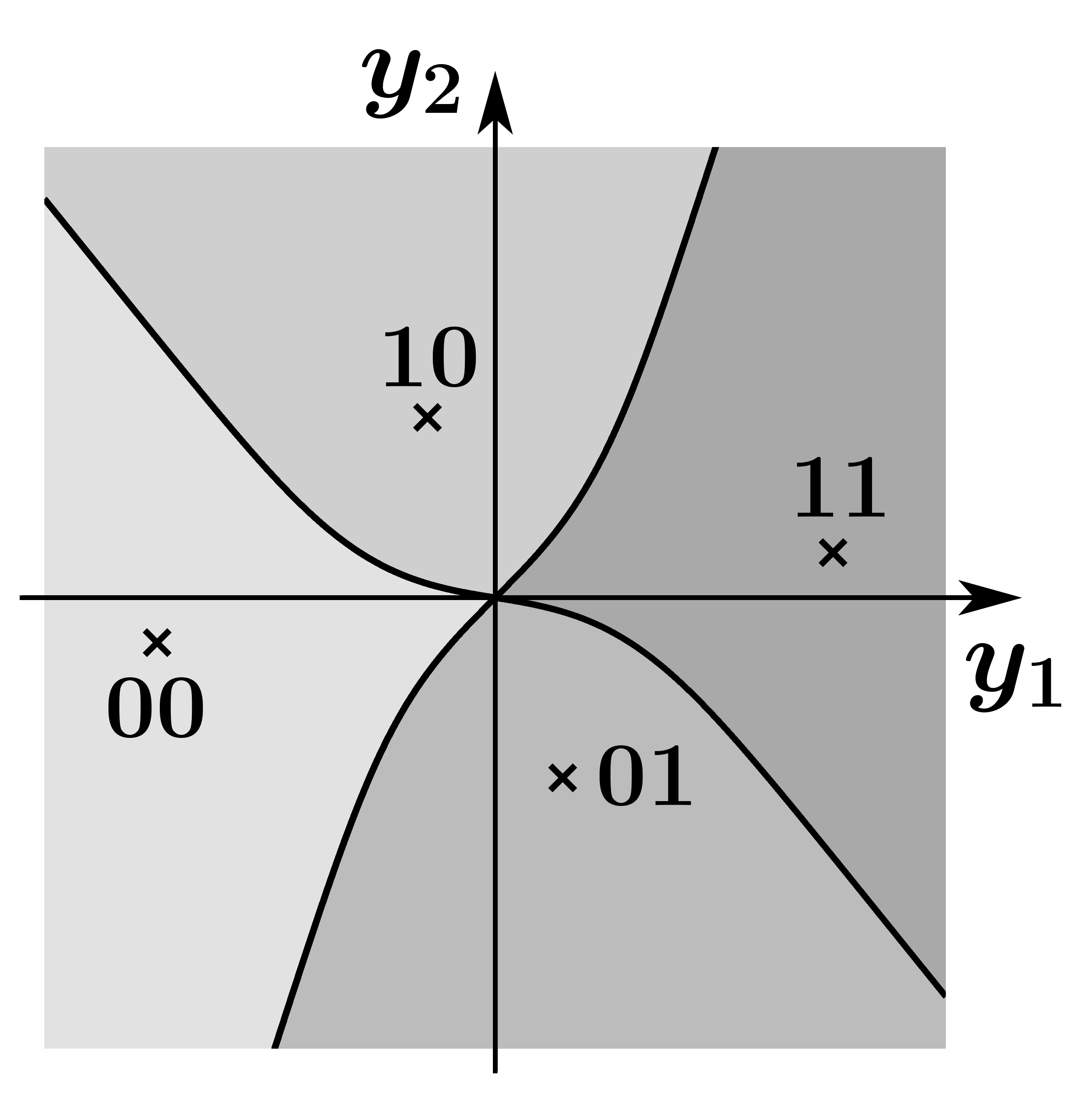}
\label{subfig:dec_bound_MMSE_5}
\vspace{-0.3cm}
}\hspace{0.4cm}
\subfloat[Vector ML]{
\centering
\includegraphics[width=0.17\textwidth]{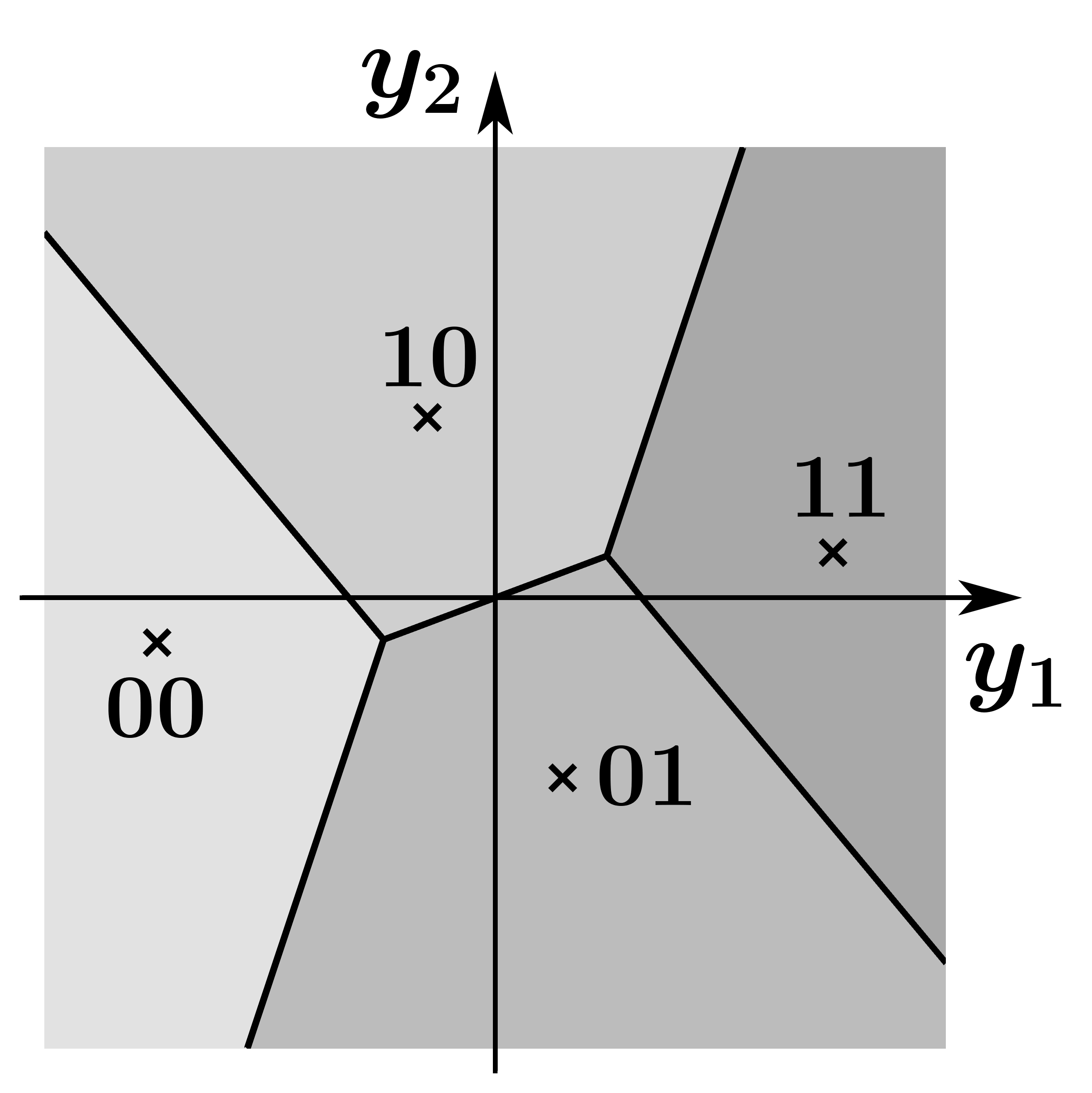}
\label{subfig:dec_bound_vecML_5}
\vspace{-0.3cm}
}\hspace{0.4cm}
\subfloat[LMMSE]{
\centering
\includegraphics[width=0.17\textwidth]{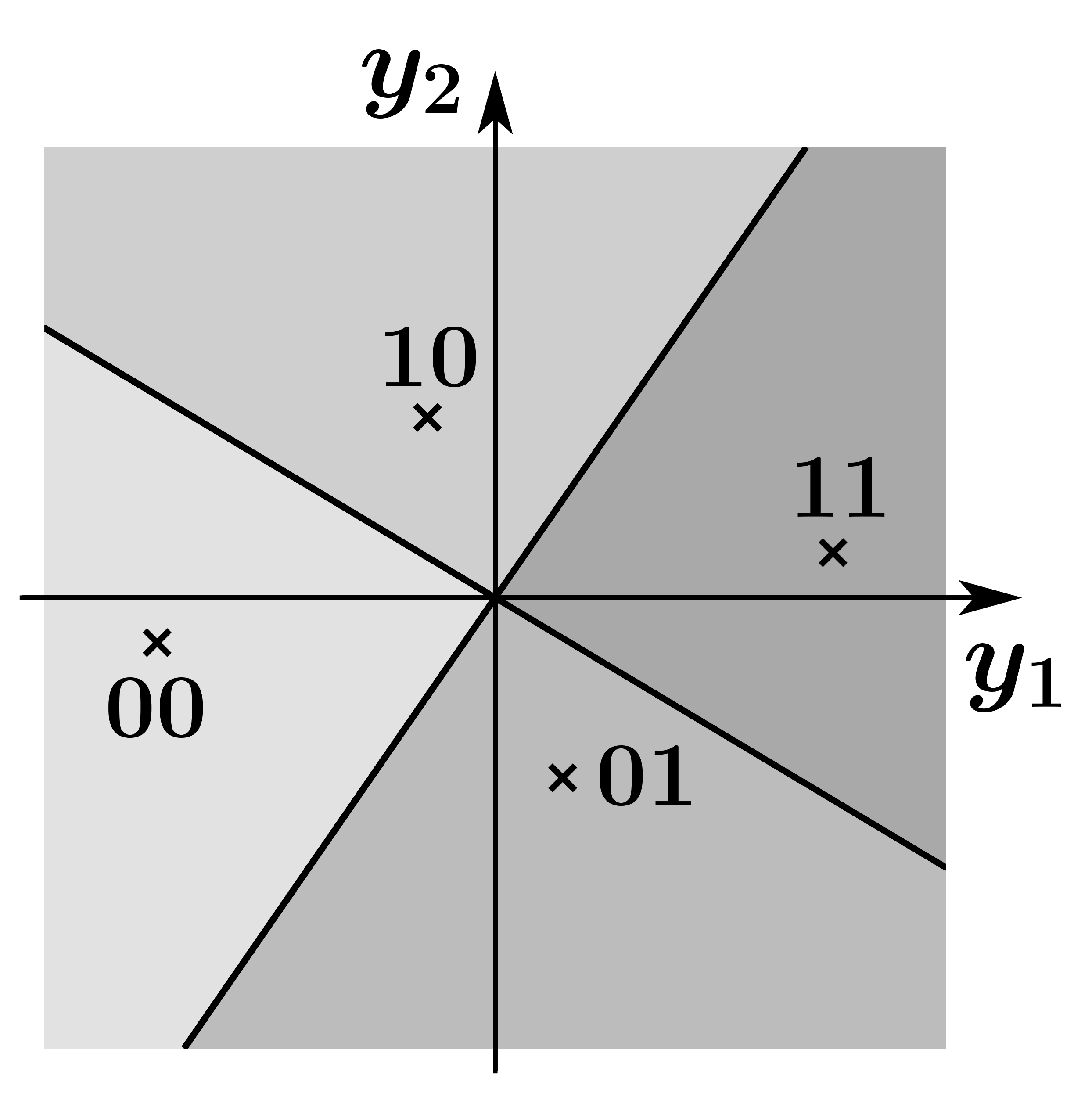}
\label{subfig:dec_bound_LMMSE_5}
\vspace{-0.3cm}
}\hspace{0.4cm}
\subfloat[DFE]{
\centering
\includegraphics[width=0.17\textwidth]{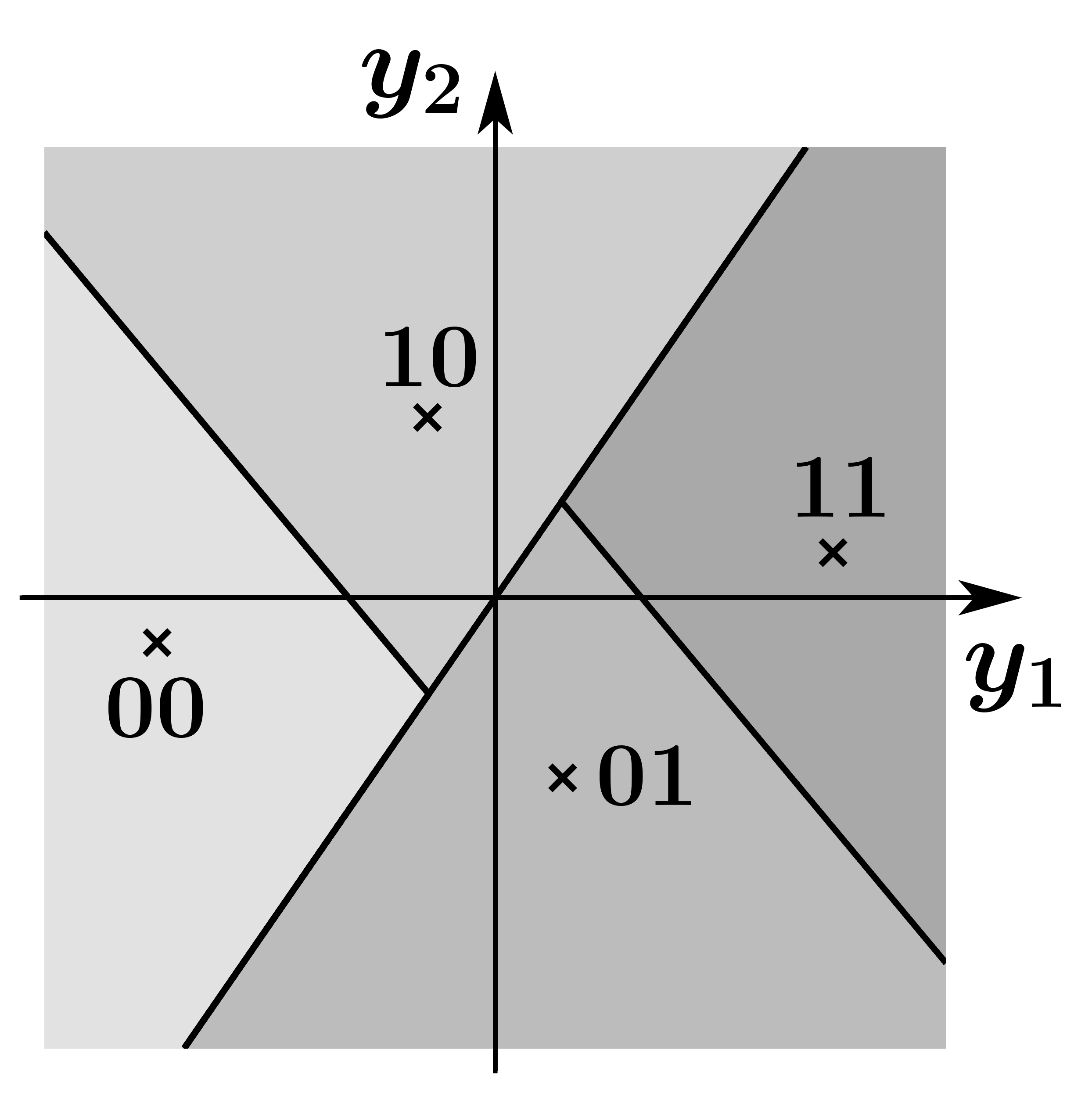}
\label{subfig:dec_bound_DFE_5}
\vspace{-0.3cm}
}
\caption{Decision boundaries of model-based equalizers for $\sigma^2=0.5$.}
\label{fig:Dec_boundaries_5dB}
\vspace{-0.3cm}
\end{figure*}
\begin{figure*}[b]
\centering
\subfloat[MMSE]{
\centering
\includegraphics[width=0.17\textwidth]{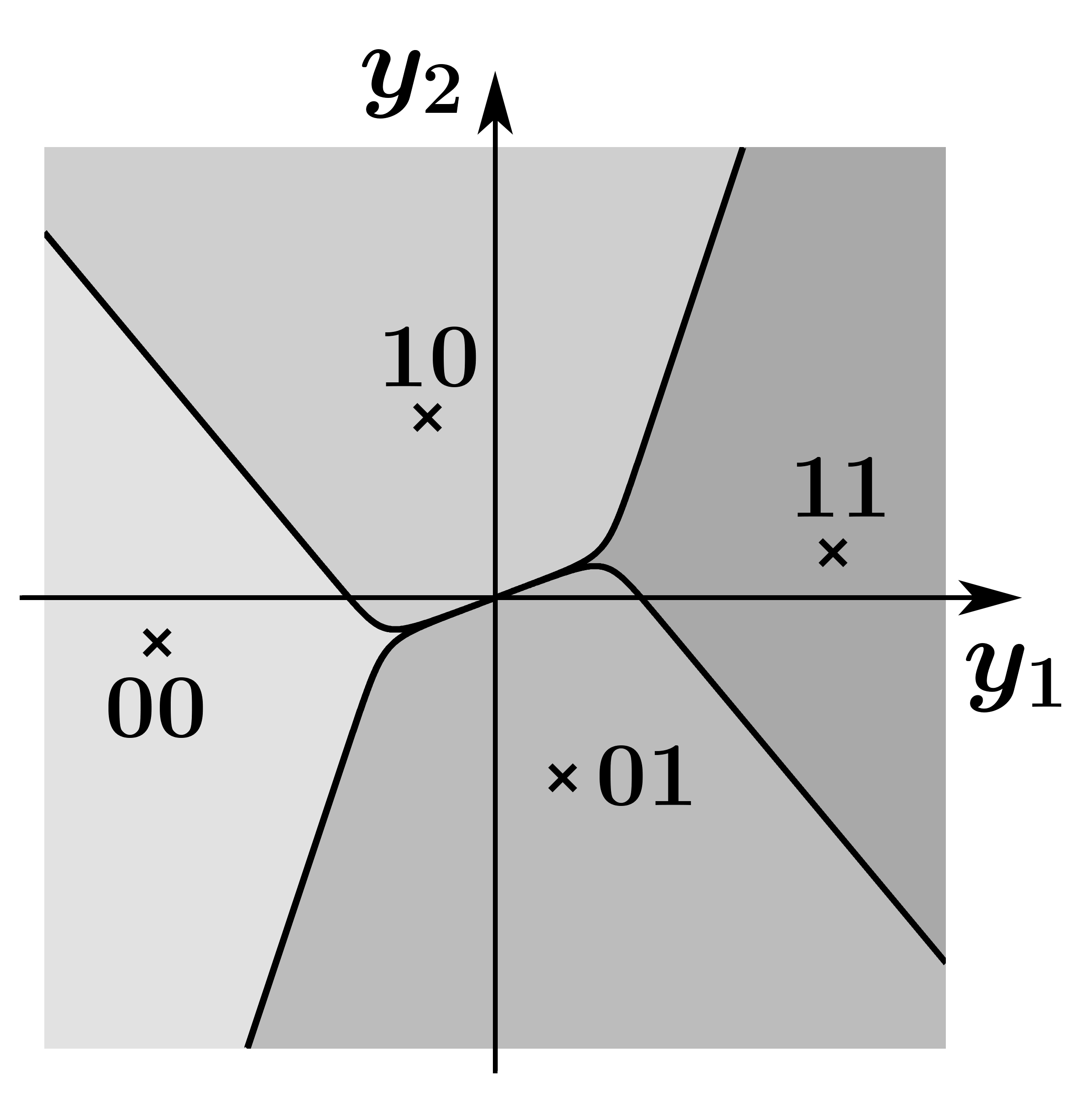}
\label{subfig:dec_bound_MMSE_15}
\vspace{-0.3cm}
}\hspace{0.4cm}
\subfloat[Vector ML]{
\centering
\includegraphics[width=0.17\textwidth]{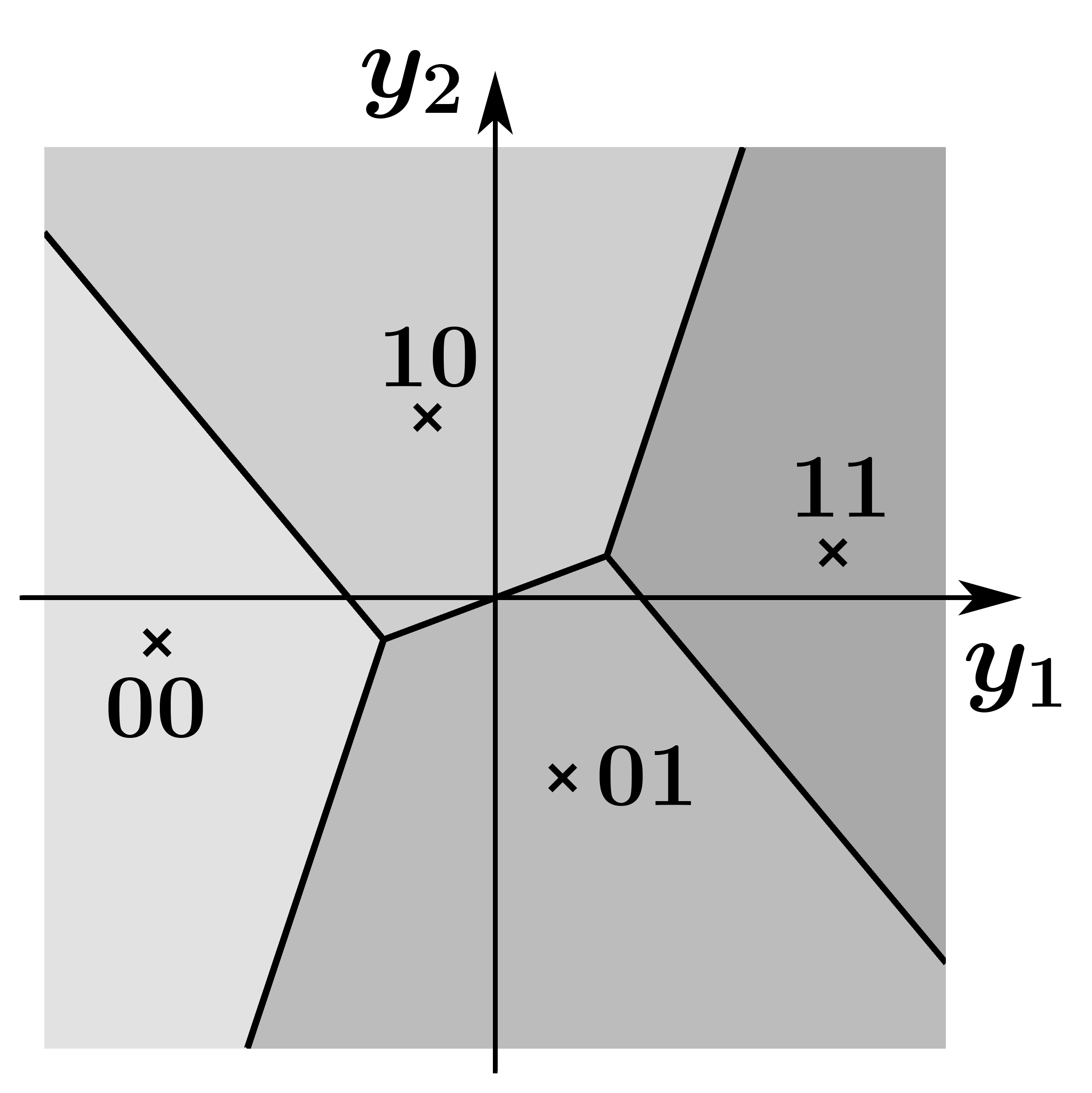}
\label{subfig:dec_bound_vecML_15}
\vspace{-0.3cm}
}\hspace{0.4cm}
\subfloat[LMMSE]{
\centering
\includegraphics[width=0.17\textwidth]{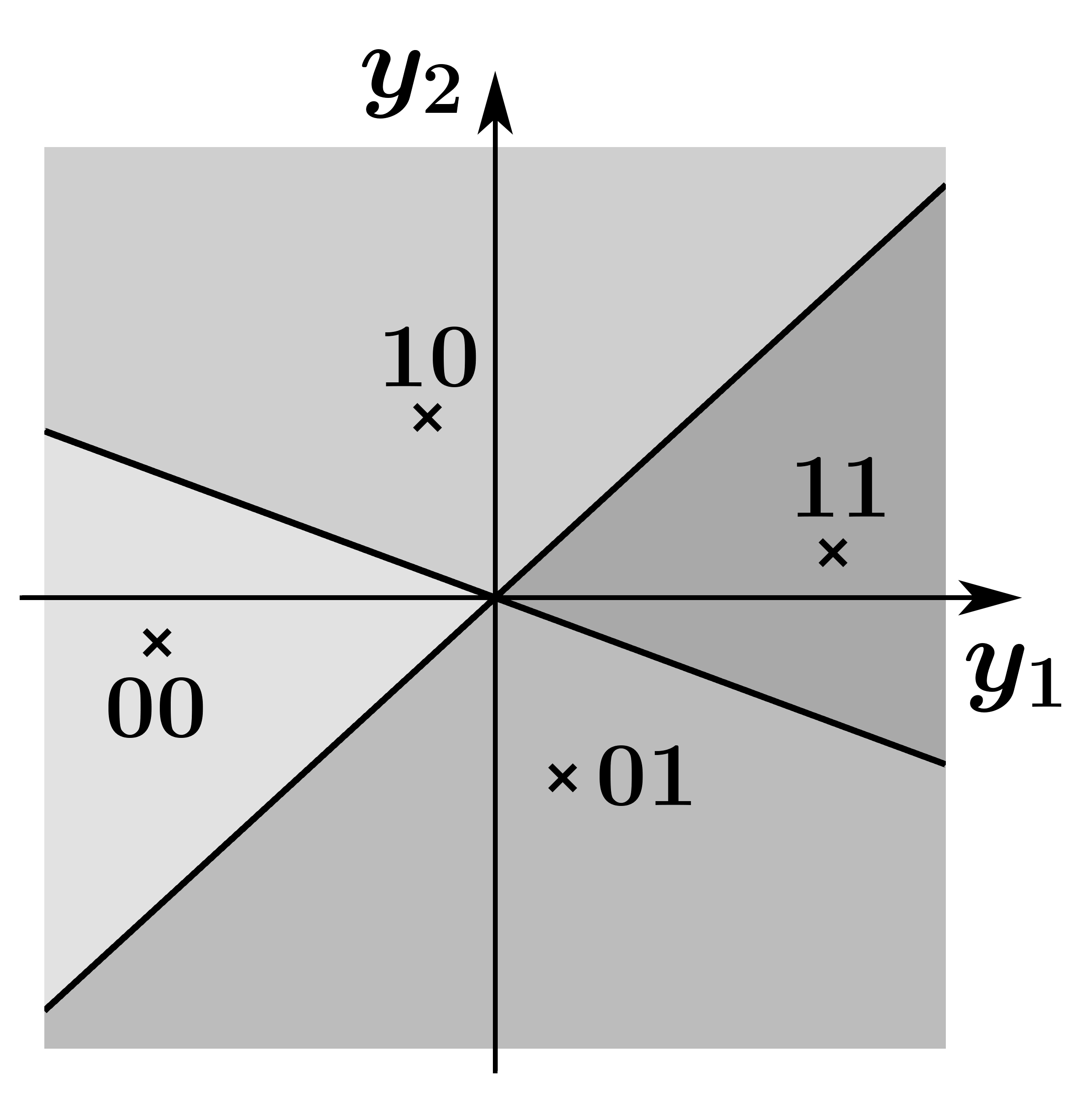}
\label{subfig:dec_bound_LMMSE_15}
\vspace{-0.3cm}
}\hspace{0.4cm}
\subfloat[DFE]{
\centering
\includegraphics[width=0.17\textwidth]{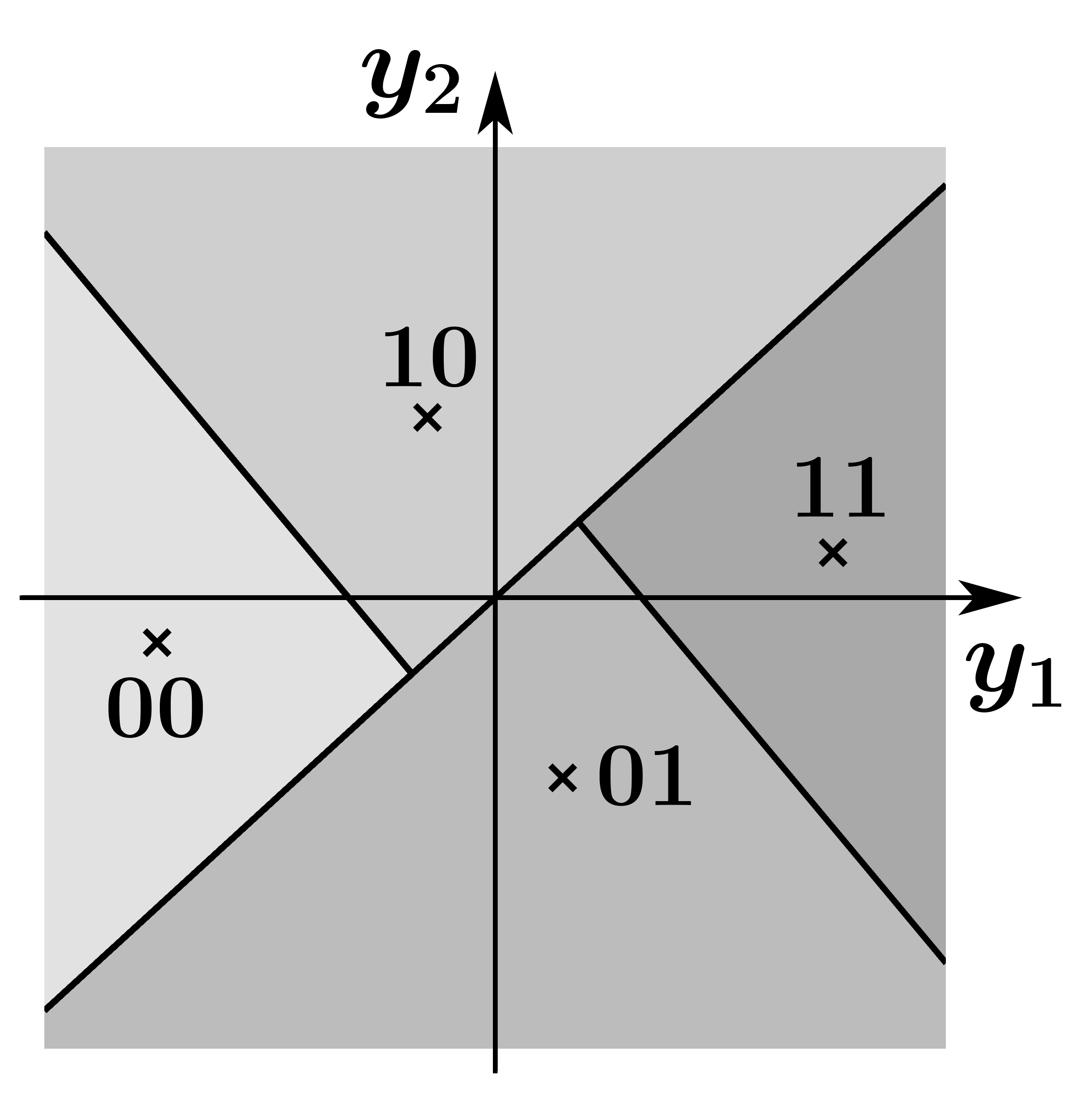}
\label{subfig:dec_bound_DFE_15}
\vspace{-0.3cm}
}
\caption{Decision boundaries of model-based equalizers for $\sigma^2=0.05$.}
\label{fig:Dec_boundaries_15dB}
\vspace{-0.4cm}
\end{figure*}

To illustrate the differences between the model-based data estimators elaborated above, we plot the decision boundaries of their hard decision estimates for a small toy example
\begin{gather}
\ve{y} = \m{H}\ve{d} + \ve{w}\,,
\end{gather}
where
\begin{gather*}
\m{H} = \begin{bmatrix}
0.9 & 0.6\\
-0.3 & 0.5
\end{bmatrix}\,,
\end{gather*}
and $\ve{w}\sim\mathcal{N}(\ve{0}, \sigma^2\m{I})$. The data vector $\ve{d} = \begin{bmatrix}
d_0 & d_1
\end{bmatrix}^T$, with $d_0, d_1 \in\{-1,1\}$, corresponds to a block of two bits $(b_1 b_0)$, where the bit values $0$ and $1$ are mapped to the symbols $-1$ and $1$, respectively. The decision boundaries are plotted for different noise power levels, i.e., for $\sigma^2 = 0.5$ and $\sigma^2=0.05$ in Fig.~\ref{fig:Dec_boundaries_5dB} and Fig.~\ref{fig:Dec_boundaries_15dB}, respectively. 

As already mentioned in Sec.~\ref{ssec:Vector_Maximum_Likelihood_Estimator}, the vector ML estimator, which is optimal regarding the estimation error probability of the whole data symbol vector, does not depend on the noise variance. This is visible in the identical decision boundaries of the vector ML estimator in Figs.~\ref{subfig:dec_bound_vecML_5} and~\ref{subfig:dec_bound_vecML_15}. The decision boundaries of the MMSE estimator (Figs.~\ref{subfig:dec_bound_MMSE_5} and~\ref{subfig:dec_bound_MMSE_15}) change with the noise variance, whereby the decision boundaries (and thus also the performance) of the MMSE estimator converge towards those of the vector ML estimator for $\sigma^2 \rightarrow 0$. That is, only for higher values of the noise variance a BER performance difference between these two equalizers might be observable. 
As shown in Sec.~\ref{ssec:Bit_Error_Ratio_Performance_Uncoded}, the BER performance difference between the vector ML estimator and the MMSE estimator is negligible for the considered UW-OFDM system.
Clearly, the decision boundaries of the LMMSE estimator (Figs.~\ref{subfig:dec_bound_LMMSE_5} and~\ref{subfig:dec_bound_LMMSE_15}), can only be straight lines. The decision boundaries of the LMMSE estimator distinctly deviate from those of the MMSE estimator, indicating a considerable performance degradation for hard decision estimation due to the linearity constraint. With the DFE, in turn, in each iteration a symbol is estimated using a linear estimation step, leading to a smaller deviation of the decision boundaries to the optimal ones, which is visible in Figs.~\ref{subfig:dec_bound_DFE_5} and~\ref{subfig:dec_bound_DFE_15}. 

\section{Neural Network Based Data Estimation}
\label{sec:Neural_Network_Based_Data_Estimation}
In this section, we introduce the utilized NN-based data estimators. We describe the actions that have to be taken specifically for UW-OFDM systems to achieve well-performing NNs, and we detail how to counteract overconfidence of NN-based data estimators to obtain reliable soft information required for channel coded data transmission. 

To use existing knowledge of NN architectures and NN training methods, real-valued input data of the NNs are generated. Hence, we map the complex-valued system model~\eqref{eq:system_model} to an equivalent real-valued model of double dimension
\begin{gather}
\ve{y} = \m{H}\ve{d} + \ve{w}\,,\label{eq:real_valued_system_model}
\end{gather}
where 
\begin{center}
\begin{minipage}{0.28\columnwidth}
\begin{gather*}
\ve{y} = \begin{bmatrix}
\text{Re}\{\ve{y}^\prime\}\\
\text{Im}\{\ve{y}^\prime\}
\end{bmatrix},
\end{gather*}
\end{minipage}%
\begin{minipage}{0.54\columnwidth}
\begin{gather*}
\m{H} = \begin{bmatrix}
\text{Re}\{\m{H}^\prime\} & -\text{Im}\{\m{H}^\prime\}\\
\text{Im}\{\m{H}^\prime\} & \text{Re}\{\m{H}^\prime\}
\end{bmatrix},
\end{gather*}
\end{minipage}
\end{center}
\begin{center}
\begin{minipage}{0.28\columnwidth}
\begin{gather*}
\ve{d} = \begin{bmatrix}
\text{Re}\{\ve{d}^\prime\}\\
\text{Im}\{\ve{d}^\prime\}
\end{bmatrix},
\end{gather*}
\end{minipage}%
\begin{minipage}{0.09\columnwidth}
\centering
\vspace{0.2cm}
and\end{minipage}%
\begin{minipage}{0.6\columnwidth}
\begin{gather*}
\ve{w} = \begin{bmatrix}
\text{Re}\{\ve{w}^\prime\}\\
\text{Im}\{\ve{w}^\prime\}
\end{bmatrix}\sim\mathcal{N}\Big(\ve{0}, \frac{N\sigma_{\text{n}}^2}{2}\m{I}\Big).
\end{gather*}
\end{minipage}
\end{center}

Assuming a symmetric alphabet $\mathbb{S}^\prime$, $\ve{d}\in\mathbb{S}^{2N_{\text{d}}}$ contains data symbols $d_i$, $i\in\{0, ..., 2N_{\text{d}}-1\}$, drawn from the real-valued symbol alphabet $\mathbb{S}~=~\text{Re}\{\mathbb{S}^\prime\}~=~\text{Im}\{\mathbb{S}^\prime\}$. The NN-based data estimators are, however, not trained to directly estimate the data symbols $d_i$, but to estimate the corresponding so-called one-hot vectors $\ve{d}_{\text{oh},i}\in\{0,1\}^{|\mathbb{S}|}$. Let $s_j\in\mathbb{S}$, $j\in\{0, ..., |\mathbb{S}|-1\}$ be the uniquely numbered symbols of the symbol alphabet $\mathbb{S}$. Then, a one-hot vector $\ve{d}_{\text{oh},i}$ corresponding to a data symbol $d_i$ that exhibits the value $s_j$ contains all zeros but a one at the $j$th position. The one-hot vectors $\ve{d}_{\text{oh},i}$ are stacked to a vector $\ve{d}_{\text{oh}}$, serving as ground truth for training the NN-based data estimators. Further, a quadratic loss function $\ell(\ve{d}_{\text{oh}}, \hat{\ve{d}}_{\text{oh}})$ is employed to quantify the error between the output $\hat{\ve{d}}_{\text{oh}}\in \mathbb{R}^{2N_{\text{d}}|\mathbb{S}|}$ of an NN and $\ve{d}_{\text{oh}}$. It can be shown (cf., e.g.,~\cite{Samuel19}), that with this approach the estimates $\hat{\ve{d}}_{\text{oh},i}$ of a properly trained NN approximately contain the posterior probabilities $\text{Pr}(d_i~=~s_j|\ve{y})$, i.e., $\hat{\ve{d}}_{\text{oh},i} \approx [\text{Pr}(d_i~=~s_0|\ve{y}),~...,~\text{Pr}(d_i~=~s_{|\mathbb{S}|-1}|\ve{y})]^T$. With the approximate posterior probabilities, LLRs can be computed using~\eqref{eq:LLR_definition}. Hence, soft information of the data symbol estimates required for coded data transmission is available. A hard decision estimate, in turn, is the symbol corresponding to the maximum entry in $\hat{\ve{d}}_{\text{oh},i}$. 

\subsection{Data Normalization}
Proper normalization of the input data of an NN is generally very important for well-behaved training, and thus the performance of trained NNs~\cite{LeCun12, Bishop95}. Interestingly, in the majority of currently available publications on NNs for data estimation in MIMO systems (e.g.,~\cite{Samuel19, He18, He20, Pratik21}), the input data of the NNs are not normalized. As we show in Sec.~\ref{ssec:Bit_Error_Ratio_Performance_Uncoded}, applying DetNet~\cite{Samuel19} as a data estimator in a UW-OFDM system without any data normalization (which is done in~\cite{Samuel19} for general MIMO systems) leads to poor BER performance. A major reason for this issue can be found by investigating the relation between the noise variance $\sigma_{\text{n}}^2$ and the signal-to-noise ratio (SNR) on receiver side. The performance of an equalizer is typically determined by evaluating the achieved BER at a specified\footnote{That is, for the evaluation of an equalizer the SNR on receiver side is specified and the noise variance is then chosen appropriately.} $E_b/N_0$, which is a measure for the SNR, where $E_b$ is the mean energy per bit, and $N_0$ is the noise power spectral density. For the following considerations, we define an SNR measure
\begin{gather}
\gamma = \frac{E_{\ve{d}}[||\m{H}\ve{d}||_2^2]}{E_{\ve{w}}[||\ve{w}||_2^2]} = \frac{\sigma_{\text{d}}^2\text{tr}(\m{H}^T\m{H})}{N^2\sigma_{\text{n}}^2}\,,
\end{gather}
which is proportional to $E_b/N_0$. For a specified SNR $\gamma$ at the input of the equalizer, the noise variance in time domain $\sigma_{\text{n}}^2$ can therefore be expressed as
\begin{gather}
\sigma_{\text{n}}^2 = \frac{\frac{1}{N}\sigma_{\text{d}}^2 \text{tr}(\m{H}^T\m{H})}{N\gamma}\,.\label{eq:noise_var_relation_SNR}
\end{gather}

In case of a general MIMO system over an uncorrelated Rayleigh fading channel, which is mainly used in, e.g., \cite{Samuel19, He18, He20, Pratik21} for the performance comparison of different NN-based data estimators, the elements of $\m{H}\in\mathbb{R}^{2N\times 2N_{\text{d}}}$ are drawn independently from a standard normal distribution, i.e., $[\m{H}]_{lm}\sim\mathcal{N}(0,1)$, $l\in\{0,~...,~2N-1\}$, $m\in\{0,~...,~2N_{\text{d}}-1\}$. This leads to $E_{\m{H}}\left[[\m{H}^T\m{H}]\right] = 2N\m{I}$.
Due to central limit theorem arguments, for large $N$, $\text{tr}(\m{H}^T\m{H})$ can be approximated as $\text{tr}(\m{H}^T\m{H}) \approx 2N_{\text{d}} E_{\m{H}}\left[[\m{H}^T\m{H}]_{ll}\right] = 4N_{\text{d}}N$. Plugging this approximation into \eqref{eq:noise_var_relation_SNR} results in
\begin{gather}
\sigma_{\text{n}}^2\approx \frac{4\sigma_{\text{d}}^2 N_{\text{d}}}{N\gamma}\,.
\end{gather}
That is, for a general MIMO system over an uncorrelated Rayleigh fading channel, the noise variances $\text{var}(w_l) = N\sigma_{\text{n}}^2$ of the elements $w_l$ of the noise vector $\ve{w}$ are independent of the current channel realization, and for a fixed SNR, they are constant. Hence, the data is implicitly normalized for this communication system. This is not the case for UW-OFDM systems, and thus we normalize the data such that the variances of the elements of the noise vector become independent of the channel realization. The data normalization is conducted by multiplying the real-valued system model~\eqref{eq:real_valued_system_model} by the normalization factor $\sqrt{N}/||\m{H}||_F$, with $||\m{H}||_F = \sqrt{\text{tr}(\m{H}^T\m{H})}$ denoting the Frobenius norm of $\m{H}$. Consequently, every element of the noise vector after normalization has a variance $\text{var}\big((\sqrt{N}w_l)/||\m{H}||_F\big) = (N^2\sigma_{\text{n}}^2)/(2||\m{H}||_F^2) = \sigma_{\text{d}}^2/(2\gamma)$, which is independent of the channel realization. This data normalization is implemented by multiplying both $\ve{y}$ and $\m{H}$ by the above-given normalization factor, which is conducted as a pre-processing step for all the NN-based data estimators presented subsequently. In the remainder of this paper, we omit the normalization factor for the sake of better readability. 

\subsection{DetNet}
\label{ssec:DetNet}
DetNet is proposed in~\cite{Samuel19} for data estimation in MIMO systems. Its network architecture is deduced by deep unfolding~\cite{Hershey14} a projected gradient descent method applied to the optimization problem of the vector ML estimator for the model~\eqref{eq:real_valued_system_model}. The $k$th step of the iterative optimization method can be expressed as 
\begin{align}
\hat{\ve{d}}_k &= \Pi\bigg(\hat{\ve{d}}_{k-1} - \left.\delta_k\frac{\partial||\ve{y} - \m{H}\ve{d}||_2^2}{\partial \ve{d}}\right|_{\ve{d}=\hat{\ve{d}}_{k-1}}\bigg)\nonumber\\
&= \Pi\big(\hat{\ve{d}}_{k-1} + 2\delta_k\m{H}^T\ve{y} - 2\delta_k\m{H}^T\m{H}\hat{\ve{d}}_{k-1}\big)\,,\label{eq:Proj_Grad_Desc_vec_ML}
\end{align} 
where $\Pi(.)$ denotes a non-linear projection to a convex subspace $\mathcal{D}$ containing all possible data vectors $\ve{d}$, i.e., $\mathbb{S}^{2N_{\text{d}}}\subset\mathcal{D}\subset\mathbb{R}^{2N_{\text{d}}}$, and $\delta_k$ is the step width in the $k$th iteration. 

The structure of the $k$th layer of the $L$ DetNet layers is inspired by a projected gradient descent iteration step~\eqref{eq:Proj_Grad_Desc_vec_ML}. Firstly, the affine mapping
\begin{gather}
\ve{q}_k = \hat{\ve{d}}_{k-1} + \delta_{k1}\m{H}^T\ve{y} - \delta_{k2}\m{H}^T\m{H}\hat{\ve{d}}_{k-1}
\label{eq:DetNet_linear_step}
\end{gather}
is applied to the layer input $\hat{\ve{d}}_{k-1}$ to obtain the temporal variable $\ve{q}_k$, where $\delta_{k1}$ and $\delta_{k2}$ are learned parameters. Secondly, the temporal variable is forwarded to a fully-connected neural network (FCNN) with a single hidden layer consisting of $d_{\text{h}}$ hidden neurons and ReLU activation, which replaces the (unknown) nonlinear projection $\Pi(.)$. To ease the training of DetNet, weighted residual connections~\cite{He16} with weighting factor $\alpha$, as well as an auxiliary loss inspired by the loss function employed for the training of GoogLeNet~\cite{Szegedy15} are utilized. Further, $d_{\text{v}}$-dimensional auxiliary variables $\ve{v}_k$ passing unconstrained information from layer to layer are used to improve the performance of DetNet. We refer to~\cite{Samuel19} for more detailed information.

\subsubsection*{Preconditioning}
Due to the deduction of the layer structure of DetNet by deep unfolding, the number of layers corresponds to the number of required iterations of the underlying projected gradient descent method. It is well known, that the condition number of the Hessian matrix in an optimization problem influences the number of iterations required for an iterative optimization method to converge. Hence, preconditioning the system model~\eqref{eq:real_valued_system_model} may reduce the number of required DetNet layers and thus the number of trainable parameters, which, in turn, enhances both the training behavior and the inference complexity. As also stated in \cite{Khani20}, we have observed~\cite{Baumgartner21_C1} that for ill-conditioned channel matrices NN-based equalizers suffer from severe performance degradation. We showed in~\cite{Baumgartner21_C1} that preconditioning distinctly narrows the eigenvalue spectrum of the Hessian matrix $\m{S}\in\mathbb{R}^{P\times P}$, $[\m{S}]_{rs} = \frac{\partial\ell(\ve{d}_{\text{oh}}, \hat{\ve{d}}_{\text{oh}})}{\partial p_r\,\partial p_s}$ of the NN learning problem, where $p_r$ and $p_s$ are two of the $P$ trainable parameters of the NN. This, in turn, allows using higher learning rates, which leads to a faster and probably better optimization of the NN parameters. We show the influence of preconditioning on the DetNet performance in Sec.~\ref{ssec:Bit_Error_Ratio_Performance_Uncoded}.

In the following, we show that preconditioning does only add a further processing step of the layer input data, while the layer structure of DetNet remains unchanged. To this end, let us rewrite the optimization problem of the vector ML estimator in form of
\begin{gather}
\min_{\ve{d}\in\mathbb{S}^{2N_{\text{d}}}}||\ve{y} - \m{H}\m{L}^{-1}\m{L}\ve{d}||_2^2\,,
\end{gather}
where $\m{L}\in\mathbb{R}^{2N_{\text{d}}\times 2N_{\text{d}}}$ is an invertible matrix. Neglecting temporarily the projection operator, a gradient descent step for the linearly transformed vector $\ve{d}_{\text{pr}} = \m{L}\ve{d}$ is given by 
\begin{align}
\hat{\ve{d}}_{\text{pr},k} &= \hat{\ve{d}}_{\text{pr},k-1} -  \left.\delta_k\frac{\partial ||\ve{y} - \m{H}\m{L}^{-1}\ve{d}_{\text{pr}}||_2^2}{\partial \ve{d}_{\text{pr}}}\right|_{\ve{d}_{\text{pr}}=\hat{\ve{d}}_{\text{pr},k-1}}\nonumber\\
&= \hat{\ve{d}}_{\text{pr},k-1}+2\delta_k\m{L}^{-T}\m{H}^T\left(\ve{y} - \m{H}\m{L}^{-1}\hat{\ve{d}}_{\text{pr},k-1}\right)\,,
\end{align}
with $\hat{\ve{d}}_{\text{pr},k/k-1} = \m{L}\hat{\ve{d}}_{k/k-1}$, and $\m{L}^{-T} = \big(\m{L} ^{-1}\big)^T = \big(\m{L}^T\big)^{-1}$. Hence, the $k$th iteration of the projected gradient descent for $\ve{d}$ follows to
\begin{align}
\hat{\ve{d}}_k &= \Pi\big(\m{L}^{-1}\hat{\ve{d}}_{\text{pr},k}\big)\nonumber\\ &= \Pi\big(\hat{\ve{d}}_{k-1} + 2\delta_k\m{P}^{-1}\m{H}^T\ve{y} - 2\delta_k\m{P}^{-1}\m{H}^T\m{H}\hat{\ve{d}}_{k-1}\big)\,,\label{eq:Precond_Proj_Grad_Desc_Step_vec_ML}
\end{align}
where $\m{P}=\m{L}^T\m{L}$ is the so-called preconditioning matrix. In this paper, we utilize a Jacobi preconditioning matrix, which is a diagonal matrix containing $\text{diag}(\m{H}^T\m{H})$ on its main diagonal. Hence, the computation of $\m{P}^{-1}$, $\m{P}^{-1}\m{H}^T\ve{y}$, and $\m{P}^{-1}\m{H}^T\m{H}$ can be carried out with low complexity. A comparison of a projected gradient descent step~\eqref{eq:Proj_Grad_Desc_vec_ML} and its preconditioned version~\eqref{eq:Precond_Proj_Grad_Desc_Step_vec_ML} reveals that preconditioning does not change the structure of the equation. Hence, the layer architecture of DetNet remains unchanged, while $\m{H}^T\ve{y}$ and $\m{H}^T\m{H}$ have to be replaced by $\m{P}^{-1}\m{H}^T\ve{y}$ and $\m{P}^{-1}\m{H}^T\m{H}$, respectively.

\subsection{Fully-Connected Neural Network}
\label{ssec:FCNN}
According to the universal approximation theorem~\cite{Hornik91}, an FCNN with a single hidden layer and sufficiently many hidden neurons can approximate any function, and thus should also be able to accomplish the task of data estimation. However, as stated in~\cite{Samuel19}, it is challenging to employ an FCNN for equalization under changing channel realizations when using the columns of $\m{H}$ concatenated with $\ve{y}$ as input data. That is, training an FCNN for different channels might be a hard task. One reason for this issue might be that no model knowledge is included in the structure of an FCNN. We therefore suggest to include model knowledge in data pre-processing.

To motivate the choice of the proposed data pre-processing with the purpose of reducing redundant information, we elucidate three observations. Firstly, the FCNN should approximate the estimator function of the optimal MMSE estimator ~\eqref{eq:MMSE_estimator}. An inspection of~\eqref{eq:MMSE_estimator} reveals, that an MMSE estimate is a sum of exponential terms, where the exponents contain\footnote{Here, we express the exponent with the real-valued quantities $\ve{y}$, $\ve{d}$, and $\m{H}$ instead of using the complex-valued $\ve{y}^\prime$, $\ve{d}^\prime$, and $\m{H}^\prime$ as in~\eqref{eq:MMSE_estimator}.}  
\begin{gather*}
||\ve{y} - \m{H}\ve{d}||_2^2 = \ve{y}^T\ve{y} - 2\ve{d}^T\m{H}^T\ve{y} + \ve{d}^T\m{H}^T\m{H}\ve{d},\quad\forall \ve{d}\in\mathbb{S}^{2 N_{\text{d}}}.
\end{gather*}
That is, the MMSE estimator does not use the isolated data $\ve{y}$ and $\ve{H}$, but the terms $\ve{y}^T\ve{y}$, $\m{H}^T\ve{y}$, and $\m{H}^T\m{H}$. Secondly, it can be shown with the help of the Fisher-Neyman factorization theorem~\cite{Casella02} that $\m{H}^T\ve{y}$ provides a sufficient statistic for the data estimation problem. Consequently, multiplying $\ve{y}$ by $\m{H}^T$, which modifies the system model to
\begin{gather}
\m{H}^T\ve{y} = \m{H}^T\m{H}\ve{d} + \m{H}^T\ve{w}\,,\label{eq:compressed_real_valued_system_model}
\end{gather}
preserves all the relevant information contained in $\ve{y}$ for the estimation of $\ve{d}$, while reducing the dimension of the available data. Thirdly, the matched filter equalizer for the system model~\eqref{eq:real_valued_system_model} is given by $\hat{\ve{d}}_{\text{MF}} = \m{H}^T\ve{y}$, which is the linear filter designed for maximizing the output SNR~\cite{Yang15}.

With the above-given arguments, we conclude that multiplying both $\m{H}$ and $\ve{y}$ by $\m{H}^T$ before using them as inputs of an FCNN compresses the input data while preserving all the information required for data estimation. Interestingly, also for DetNet the quantities $\m{H}^T\ve{y}$ and $\m{H}^T\m{H}$ are utilized instead of $\ve{y}$ and $\m{H}$, however, due to a different motivation, and in a different manner. Since $\m{H}^T\m{H}$ is a symmetric matrix, the dimension of the input data is further reduced by utilizing only the upper triangular matrix of $\m{H}^T\m{H}$ including its main diagonal. That is, the input vector of the FCNN data estimator is
\begin{align*}
\big[ [\m{H}^T\m{H}]_{00},\,&[\m{H}^T\m{H}]_{0:1,1}^T,\,[\m{H}^T\m{H}]_{0:2,2}^T,\,\cdots, \\ &[\m{H}^T\m{H}]_{0:2N_{\text{d}}-1,2N_{\text{d}}-1}^T,\,(\m{H}^T\ve{y})^T\big]^T,
\end{align*}
where $[\m{H}^T\m{H}]_{0:l, l}$ denotes the vector containing the first $l+1$ entries of the $l$th column of $\m{H}^T\m{H}$.

The utilized FCNNs for equalization are comprised of $L$ layers, $d_{\text{h}}$ neurons per hidden layer, and weighted residual connections with weighting factor $\alpha$. The employed activation functions $\varphi(.)$ are stated in Tab.~\ref{tab:Hyperparameter_settings}.
\subsection{Attention Detector}
\begin{figure}[t]
\centering
\begin{tikzpicture}[scale=.75, every node/.style={transform shape}]
\tikzset{
ddots/.pic={
\draw[fill] (0,0) circle [radius=0.03];
\draw[fill] (-0.3,0) circle [radius=0.03];
\draw[fill] (0.3,0) circle [radius=0.03];
}
}
\coordinate (A) at (0,0);
\coordinate (B) at ($(A) + (-3,0)$);
\coordinate (C) at ($(A) + (3,0)$);
\node [draw, minimum width=3cm] at ($(B) + (0,1)$) (add_norm1) {Add \& Norm};
\node [draw, minimum width=3cm] at ($(C) + (0,1)$) (add_norm2) {Add \& Norm};
\node at ($(B) + (0,-1.3)$) (inp1) {$\ve{m}_0^{(l)\,T}$};
\node at ($(C) + (0,-1.3)$) (inpN) {$\ve{m}_{N_{\text{d}}-1}^{(l)\,T}$};
\draw[->,rounded corners] ($(inpN)+(0,0.6)$) -- ++(-2,0) |- (add_norm2);
\node [draw, minimum width=9cm, fill=white] at (A) (attention_rect) {Self-Attention Layer};
\draw[->] (attention_rect.north -| add_norm1) -- (add_norm1);
\draw[->] (attention_rect.north -| add_norm2) -- (add_norm2);
\draw[->] (inp1.north) -- (inp1 |- attention_rect.south);
\draw[->] (inpN.north) -- (inpN |- attention_rect.south);
\node [draw, minimum width=3cm] at ($(add_norm1) + (0,1)$) (ff1) {Feedforward};
\node [draw, minimum width=3cm] at ($(add_norm2) + (0,1)$) (ff2) {Feedforward};
\draw[->] (add_norm1) -- (ff1);
\draw[->] (add_norm2) -- (ff2);
\node [draw, minimum width=3cm] at ($(ff1) + (0,1)$) (add_norm3) {Add \& Norm};
\node [draw, minimum width=3cm] at ($(ff2) + (0,1)$) (add_norm4) {Add \& Norm};
\draw[->] (ff1) -- (add_norm3);
\draw[->] (ff2) -- (add_norm4);
\draw[->,rounded corners] ($(inp1)+(0,0.6)$) -- ++(-2,0) |- (add_norm1);
\draw[->,rounded corners] ($(add_norm1)+(0,0.45)$) -- ++(-2,0) |- (add_norm3);
\draw[->,rounded corners] ($(add_norm2)+(0,0.45)$) -- ++(-2,0) |- (add_norm4);
\pic at ($(inp1)!0.5!(inpN)$) {ddots};
\pic at ($(add_norm1)!0.5!(add_norm2)$) {ddots};
\pic at ($(ff1)!0.5!(ff2)$) {ddots};
\pic at ($(add_norm3)!0.5!(add_norm4)$) {ddots};
\node at ($(add_norm3) + (0,1)$) (out1) {$\ve{m}_0^{(l+1)\,T}$};
\node at ($(add_norm4) + (0,1)$) (outN) {$\ve{m}_{N_{\text{d}}-1}^{(l+1)\,T}$};
\draw [->] (add_norm3) -- (out1);
\draw [->] (add_norm4) -- (outN);
\end{tikzpicture}
\caption{Structure of one encoder layer of the Attention Detector.}
\label{fig:Structure_encoder_layer_Attention_Detector}
\end{figure}
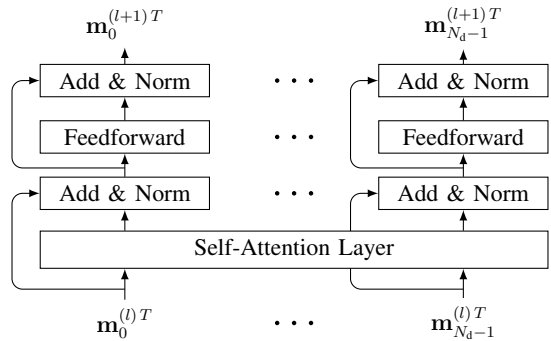

Due to the arguments given in Sec.~\ref{ssec:FCNN}, we use the compressed system model~\eqref{eq:compressed_real_valued_system_model} for defining the inputs of the so-called Attention Detector. Investigations on the compressed system model~\eqref{eq:compressed_real_valued_system_model} revealed that the entries in $\m{H}^T\ve{y}$ are correlated. This observation motivates the use of the self-attention mechanism~\cite{Vaswani17} to exploit these correlations for enhancing the estimation performance and/or reducing the required computational complexity. For further elaborations on the network architecture of the Attention Detector, let us start by defining its inputs, which are the rows $\ve{m}_k^T$, $k\in\{0, ..., 2N_{\text{d}}-1\}$, of the matrix
\begin{gather}
\m{M} = \m{P}^{-1}\big[\m{H}^T\ve{y},\, \m{H}^T\m{H}\big]\,,\label{eq:input_Attention_detector}
\end{gather}
where $\m{P}$ is the Jacobi preconditioning matrix as described in Sec.~\ref{ssec:DetNet}. Although the layer architecture of the Attention Detector is not deduced by deep unfolding, we apply preconditioning for obtaining a narrower eigenvalue spectrum of the Hessian matrix of the NN learning problem, cf. Sec.~\ref{ssec:DetNet}.
The vectors $\ve{m}_k^{T}$ serve as an input sequence of an encoder. Since the rows of the equation system~\eqref{eq:compressed_real_valued_system_model} are interchangeable, no positional encoding is applied to the vectors. The encoder is very similar to that of the Transformer~\cite{Vaswani17}. It is comprised of $L_{\text{enc}}$ stacked encoder layers, whereby the $l$th encoder layer, $l\in\{0, ..., L_{\text{enc}}-1\}$, is schematically shown in Fig.~\ref{fig:Structure_encoder_layer_Attention_Detector}. An encoder layer with inputs\footnote{Note that $\ve{m}_k^{(0)\,T} = \ve{m}_k^T$.} $\ve{m}_k^{(l)\,T}$ consists of a self-attention layer~\cite{Vaswani17}, followed by a batch norm layer, a single hidden layer FCNN with $d_{\text{h,enc}}$ hidden neurons and ReLU activation function, and another batch norm layer. Around both the self-attention layer and the single hidden layer FCNN residual connections are employed. Further, dropout~\cite{Srivastava14} with a dropout rate $D$ is applied to the outputs of the self-attention layer and the single hidden layer FCNN, as well as to the input layer outputs of the latter.  The outputs of the last encoder layer $\ve{m}_{k}^{(L_{\text{enc}}-1)\,T}$ are concatenated to the input vector $\ve{s} = [\ve{m}_{0}^{(L_{\text{enc}}-1)\,T},\,\cdots,\,\ve{m}_{N_{\text{d}}-1}^{(L_{\text{enc}}-1)\,T}]$ of a shallow FCNN with $L_{\text{fcnn}}$ hidden layers, $d_{\text{h,fcnn}}$ neurons per hidden layer, and an activation function $\varphi(.)$ specified in Tab.~\ref{tab:Hyperparameter_settings}. The outputs of this shallow FCNN are the final estimation results $\hat{\ve{d}}_{\text{oh}}$. 

\section{Results}
\label{sec:Results}
In this section, we compare the presented NN-based data estimators with state-of-the-art model-based equalizers in terms of the achieved BER performance over a specified SNR range by means of simulations. We study the equalizers for channel coded and uncoded data transmission, and we regard their computational complexity. Further, we highlight the peculiar distribution of the estimates provided by NN-based data estimators. Due to the multitude of possible combinations of system settings, only selected simulation cases are presented, while those setups that do not provide further insights are omitted. Since with non-systematic UW-OFDM signaling a better BER performance is achievable~\cite{HuemHofbHub12}, we focus on this signaling scheme in our investigations. 

\subsection{Simulation Setup}
\label{ssec:Simulation_Setup}
The evaluation is conducted for two different system dimensions. The parameter setup for system~I is $N=12$, $N_{\text{d}} = 8$, $N_{\text{u}} = 4$, $N_{\text{z}} = 0$, and $N_{\text{p}} = 0$, and for system~II $N=64$, $N_{\text{d}} = 32$, $N_{\text{u}}=16$, $N_{\text{z}}=12$, and $N_\text{p}=4$.  System~II should represent a real-world communication system, where $N_{\text{p}}$ pilot subcarriers can be utilized for synchronization purposes. However, in our simulations, the pilot subcarriers are unused and do not influence the presented results. For system~I, in turn, the BER performance of the optimal model-based data estimators can be simulated in a reasonable time, which allows providing insights concerning the gap between the performance achieved with an NN-based equalizer and the lower BER bound.

We assume data transmission over a multipath channel in form of data bursts comprised of a sequence of $1000$ UW-OFDM symbols. The channel is assumed to be stationary for a single data burst, but to be changing independently of all other channel realizations from burst to burst. We utilize the statistical channel model~\cite{Fakatselis97} of an indoor frequency selective environment, where the channel impulse responses are modeled in form of tapped delay lines. The complex tap values exhibit a uniformly distributed phase and a Rayleigh distributed magnitude with an exponentially decaying power profile. As in the referenced works on UW-OFDM~\cite{HuemHofbHub10, HuemHofbHub12,  Hofbauer16}, we use for system~II a sampling time $T_\text{s}=50\,\si{ns}$, and we choose a channel delay spread of $\tau_{\text{RMS}}=100\,\si{ns}$. For system~I, we specify the sampling time to be $200\,\si{ns}$ while keeping the same channel delay spread as for system~II. We assume perfect channel knowledge on receiver side. The presented BER curves are obtained by averaging over $8000$ channels. 

For channel coded data transmission, a convolutional code with generator polynomials $(133, 171)_8$, constraint length $7$, and rate $R=1/2$ is used, whereby a Viterbi channel decoder is employed. As already mentioned, the data symbols are drawn from a QPSK modulation alphabet.

\subsection{Neural Network Training}
\label{ssec:NN_Training}
{
\renewcommand{\aboverulesep}{0pt}
\renewcommand{\belowrulesep}{0pt}
\renewcommand{\arraystretch}{1.3}
\begin{table*}[t]
\caption{Hyperparameter settings.}
\label{tab:Hyperparameter_settings}
\begin{center}
\begin{tabularx}{\textwidth}{>{\raggedright}m{1.1cm}  |>{\raggedleft}m{1.2cm} >{\raggedleft}X >{\raggedleft}m{0.6cm} >{\raggedleft}X >{\raggedleft}X | >{\raggedleft}m{1.2cm} >{\raggedleft}X >{\raggedleft}X >{\raggedleft}X >{\raggedleft}m{0.6cm} | >{\raggedleft}m{1.2cm} >{\raggedleft}X >{\raggedleft}m{0.7cm} >{\raggedleft}m{0.6cm} >{\raggedleft}m{0.7cm} >{\raggedleft}X >{\raggedleft\arraybackslash}m{0.6cm}}
\toprule
& \multicolumn{5}{c|}{DetNet} &\multicolumn{5}{c|}{FCNN} & \multicolumn{7}{c}{Attention Detector}\\
& \centering $\eta$ & \centering $L$ & \centering $d_{\text{h}}$ & \centering $d_{\text{v}}$ & \centering $\alpha$ &  \centering $\eta$ & \centering $L$ & \centering $d_{\text{h}}$ & \centering $\alpha$ &  \centering $\varphi(.)$ &\centering $\eta$ & \centering $L_{\text{enc}}$ & \centering $d_{\text{h,enc}}$ & \centering $L_{\text{fcnn}}$ & \centering $d_{\text{h,fcnn}}$ & \centering $D$ & \centering $\varphi(.)$ \tabularnewline
\hline
System~I & $1.9\cdot 10^{-3}$ & $10$ & $80$ & $32$ & $0.1$ & $6.0\cdot 10^{-4}$ & $10$ & $300$ & $0.0$ & ReLU &$1.8\cdot 10^{-3}$ & $8$ & $80$ & $2$ & $150$ & $0.0$ & ReLU\\
\hline
System~II &$4.6\cdot 10^{-4}$ & $30$& $250$& $80$ & $0.9$ & $1.0\cdot 10^{-4}$ & $22$ & $800$ & $0.7$ & SeLU &$3.0\cdot 10^{-4}$  & $10$ & $400$ & $3$ & $500$ &$0.0$ & SeLU\\
\bottomrule
\end{tabularx}
\end{center}
\end{table*}
}

The dataset for training the NNs is obtained by simulating sample data transmissions with known payload data over randomly generated multipath channels following the employed channel model described in Sec.~\ref{ssec:Simulation_Setup}. Since data estimation is most challenging for transmissions over deep fading channels, we emphasize those cases by adding a set of sample data transmissions to the training set that solely contains transmissions over deep fading channels. The channels for this subset of the training set are found by creating $5000$ times more channels than needed and picking the channels with the most severe fading holes. Including particularly bad channels in the training set turns out to be beneficial for the BER performance of the NN-based data estimators (a similar observation has also been mentioned in~\cite{Wiesel18}). Empirical investigations show that the proportion of the subset of specifically generated bad channels in the training set of $10\,\%$ and $50\,\%$ is a good choice for system~I and system~II, respectively. Overall, the training set consists of $30000$ channels and $40000$ channels for system~I and system~II, respectively. The selection of the $E_b/N_0$ values for the sample data transmissions, which turns out to have a major impact on the performance of the NNs, differs for the simulated system setups, and thus is given with the results for the chosen system setup. 
Furthermore, we pre-trained the NNs with noiseless data transmissions, i.e., the sent data is only disturbed by a multipath channel, over $2000$ different channels, which leads to a faster training convergence.

For the training, we employ an Adam optimizer~\cite{Kingma14} with default settings. The learning rate is decreased exponentially, such that the learning rate in the final optimization step is $5\,\%$ of the initial learning rate $\eta$. All NNs are trained with a batch size of $1024$ and for $60$ epochs. Further, early stopping is utilized as a regularization technique. The hyperparameters of the NN-based equalizers are found with an extensive grid search by evaluating the trained NNs on a validation set; the best settings found are summarized in Tab.~\ref{tab:Hyperparameter_settings}.

\subsection{Bit Error Ratio Performance -- Uncoded Transmission}
\label{ssec:Bit_Error_Ratio_Performance_Uncoded}

\begin{figure}[t]
\begin{center}
\begin{tikzpicture}[
    spy/.style={%
        draw,black,
        line width=0.4pt,
        circle,inner sep=0pt,
    },
]
    \def\spyviewersize{2.6cm}

    \def\spyonclipreduce{0.4pt}

    \def\spyfactorI{15}
    \coordinate (spy-in 1) at (1.35,1.4);
    \coordinate (spy-on 1) at (0.93,3.67);

\def\pik{
\begin{semilogyaxis}[compat=newest, width=6.5cm, height=6.5cm, grid=both, ylabel={\small BER}, ymax = 1e-2, ymin = 1e-6, xmax=16, xmin=8, xlabel={\small $E_b/N_0$ (dB)}, scaled ticks = false,  x tick label style={/pgf/number format/.cd,fixed,precision=2,/tikz/.cd}, ticklabel style={font=\scriptsize}, legend style={at={(0.36,1.03)}, anchor=south}, legend columns = {3}, legend cell align=left, legend style={font=\scriptsize}, every axis plot/.append style={thick}]
\addplot[color=myblue, mark=o, smooth, mark size=0.05cm] table[col sep=semicolon, x=EbN0_dB, y=ber_DetNet]{./fig/20210317113627_small_non_sys_uncoded_QPSK/BER_estimators.csv};
\addlegendentry{\scriptsize DetNet  w/ norm. w/ precond.}
\addplot[color=myorange, mark=x, smooth] table[col sep=semicolon, x=EbN0_dB, y=ber_AttentionDetector]{./fig/20210317113627_small_non_sys_uncoded_QPSK/BER_estimators_comprInpFCNN_AttDet.csv};
\addlegendentry{\scriptsize Attention Detector}
\addplot[color=mygreen, mark=triangle, smooth] table[col sep=semicolon, x=EbN0_dB, y=ber_DF]{./fig/20210317113627_small_non_sys_uncoded_QPSK/BER_estimators.csv};
\addlegendentry{\scriptsize DFE}
\addplot[color=myblue, mark=o, dashed, mark options={solid}, smooth, mark size=0.05cm] table[col sep=semicolon, x=EbN0_dB, y=ber_DetNet]{fig/20210112212129_influence_preprocessing/BER_estimators_wo_precond_latest.csv};
\addlegendentry{\scriptsize DetNet w/ norm. w/o precond.}
\addplot[color=myviolet, mark=+, smooth] table[col sep=semicolon, x=EbN0_dB, y=ber_FCNNCompressedInp]{./fig/20210317113627_small_non_sys_uncoded_QPSK/BER_estimators_comprInpFCNN_AttDet.csv};
\addlegendentry{\scriptsize FCNN}
\addplot[color=black, mark=square, smooth] table[col sep=semicolon, x=Eb_N0_dB, y=ber_mmse]{./fig/20210317113627_small_non_sys_uncoded_QPSK/Matlab_results_MMSE.csv};
\addlegendentry{\scriptsize MMSE}
\addplot[color=myblue, mark=o, densely dotted, mark options={solid}, smooth, mark size=0.05cm] table[col sep=semicolon, x=EbN0_dB, y=ber_DetNet]{fig/20210112212129_influence_preprocessing/BER_estimators_wo_normalization_wo_precond.csv};
\addlegendentry{\scriptsize DetNet w/o norm. w/o precond.}
\addplot[color=myred, mark=diamond, smooth] table[col sep=semicolon, x=EbN0_dB, y=ber_lin_est]{./fig/20210317113627_small_non_sys_uncoded_QPSK/BER_estimators.csv};
\addlegendentry{\scriptsize LMMSE}
\addplot[color=mykaki, mark=+, smooth] table[col sep=semicolon, x=Eb_N0_dB, y=ber_mle]{./fig/20210317113627_small_non_sys_uncoded_QPSK/Matlab_results_ML_est.csv};
\addlegendentry{\scriptsize Vector ML}

\end{semilogyaxis}}
\pik

\def\pik2{
\begin{semilogyaxis}[compat=newest, width=6.5cm, height=6.5cm, grid=both, minor grid style={line width=0.05pt}, major grid style={line width=0.05pt}, ylabel={\small BER}, ymax = 1e-2, ymin = 1e-6, xmax=16, xmin=8, xlabel={\small $E_b/N_0$ (dB)}, scaled ticks = false,  x tick label style={/pgf/number format/.cd,fixed,precision=2,/tikz/.cd}, ticklabel style={font=\scriptsize}, every axis plot/.append style={line width=0.1pt}]
\addplot[color=myblue, mark=o, smooth, mark size=0.05cm] table[col sep=semicolon, x=EbN0_dB, y=ber_DetNet]{./fig/20210317113627_small_non_sys_uncoded_QPSK/BER_estimators.csv};
\addplot[color=myorange, mark=x, smooth] table[col sep=semicolon, x=EbN0_dB, y=ber_AttentionDetector]{./fig/20210317113627_small_non_sys_uncoded_QPSK/BER_estimators_comprInpFCNN_AttDet.csv};
\addplot[color=mygreen, mark=triangle, smooth] table[col sep=semicolon, x=EbN0_dB, y=ber_DF]{./fig/20210317113627_small_non_sys_uncoded_QPSK/BER_estimators.csv};
\addplot[color=myblue, mark=o, dashed, mark options={solid}, smooth, mark size=0.05cm] table[col sep=semicolon, x=EbN0_dB, y=ber_DetNet]{fig/20210112212129_influence_preprocessing/BER_estimators_wo_precond_latest.csv};
\addplot[color=myviolet, mark=+, smooth] table[col sep=semicolon, x=EbN0_dB, y=ber_FCNNCompressedInp]{./fig/20210317113627_small_non_sys_uncoded_QPSK/BER_estimators_comprInpFCNN_AttDet.csv};
\addplot[color=black, mark=square, smooth] table[col sep=semicolon, x=Eb_N0_dB, y=ber_mmse]{./fig/20210317113627_small_non_sys_uncoded_QPSK/Matlab_results_MMSE.csv};
\addplot[color=myblue, mark=o, densely dotted, mark options={solid}, smooth, mark size=0.05cm] table[col sep=semicolon, x=EbN0_dB, y=ber_DetNet]{fig/20210112212129_influence_preprocessing/BER_estimators_wo_normalization_wo_precond.csv};
\addlegendentry{\scriptsize DetNet w/o norm. w/o precond.}
\addplot[color=myred, mark=diamond, smooth] table[col sep=semicolon, x=EbN0_dB, y=ber_lin_est]{./fig/20210317113627_small_non_sys_uncoded_QPSK/BER_estimators.csv};
\addplot[color=mykaki, mark=+, smooth] table[col sep=semicolon, x=Eb_N0_dB, y=ber_mle]{./fig/20210317113627_small_non_sys_uncoded_QPSK/Matlab_results_ML_est.csv};

\end{semilogyaxis}}

    \node[spy,minimum size={\spyviewersize/\spyfactorI}] (spy-on node 1) at (spy-on 1) {};
    \node[spy,minimum size=\spyviewersize] (spy-in node 1) at (spy-in 1) {};
    \begin{scope}
        \clip (spy-in 1) circle (0.5*\spyviewersize-\spyonclipreduce);
        \node[circle, fill=white, draw=white, minimum size=\spyviewersize] (background_circ) at (spy-in 1) {};
        \pgfmathsetmacro\sI{1/\spyfactorI}
        \begin{scope}[
            shift={($\sI*(spy-in 1)-\sI*(spy-on 1)$)},
            scale around={\spyfactorI:(spy-on 1)}
        ]
            \pik2
        \end{scope}
    \end{scope}
    \draw [spy] (spy-on node 1) -- (spy-in node 1);

\end{tikzpicture}
\vspace{-0.3cm}
\caption{BER performance comparison for system I, uncoded case.}
\label{fig:system_I_non_sys_uncoded_QPSK}
\end{center}
\end{figure}
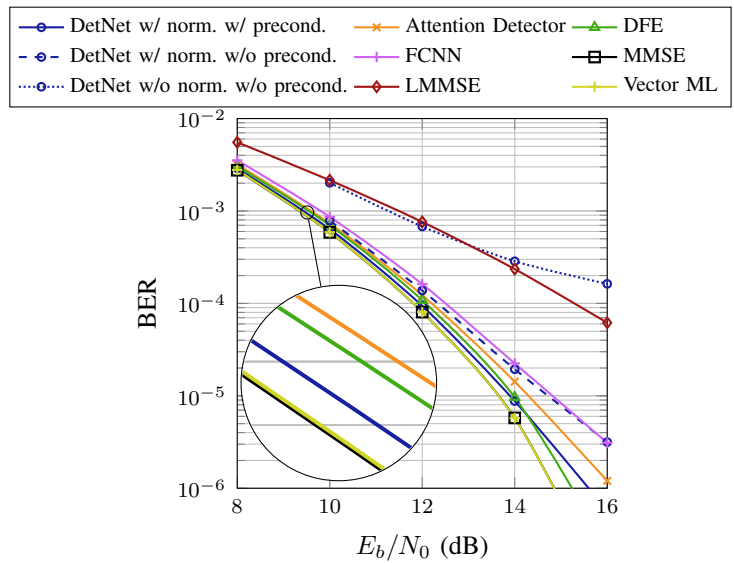

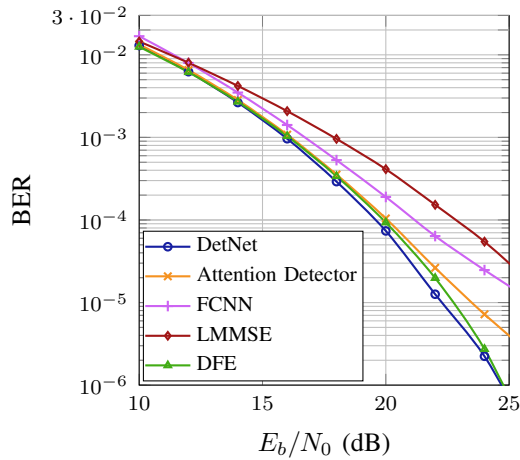
\begin{figure}[t]
\begin{center}
\begin{tikzpicture}
\begin{semilogyaxis}[compat=newest, width=6.5cm, height=6.5cm, grid=both, ylabel={\small BER}, ymax = 3e-2, ymin = 1e-6, xmax=25, xmin=10, xlabel={\small $E_b/N_0$ (dB)}, scaled ticks = false,  x tick label style={/pgf/number format/.cd,fixed,precision=2,/tikz/.cd}, ticklabel style={font=\scriptsize}, extra y ticks={2e-2, 3e-2}, extra y tick labels={,$3\cdot 10^{-2}$}, legend columns = {1}, legend cell align=left, legend style={font=\scriptsize, inner sep = 0.3pt, at={(0.01,0.01)}, anchor=south west}, every axis plot/.append style={thick}]
\addplot[color=myblue, mark=o, smooth, mark size=0.05cm] table[col sep=semicolon, x=EbN0_dB, y=ber_DetNet]{./fig/20220912170545_large_non_sys_uncoded_QPSK/BER_estimators_DetNet.csv};
\addlegendentry{\scriptsize DetNet}
\addplot[color=myorange, mark=x, smooth] table[col sep=semicolon, x=EbN0_dB, y=ber_AttentionDetector]{./fig/20220912170545_large_non_sys_uncoded_QPSK/BER_estimators_AttDet_FCNN.csv};
\addlegendentry{\scriptsize Attention Detector}
\addplot[color=myviolet, mark=+, smooth] table[col sep=semicolon, x=EbN0_dB, y=ber_FCNNCompressedInpResCon]{./fig/20220912170545_large_non_sys_uncoded_QPSK/BER_estimators_AttDet_FCNN.csv};
\addlegendentry{\scriptsize FCNN}
\addplot[color=myred, mark=diamond, smooth, mark size=0.05cm] table[col sep=semicolon, x=EbN0_dB, y=ber_lin_est]{./fig/20220912170545_large_non_sys_uncoded_QPSK/BER_estimators_DetNet.csv};
\addlegendentry{\scriptsize LMMSE}
\addplot[color=mygreen, mark=triangle, smooth, mark size=0.05cm] table[col sep=semicolon, x=EbN0_dB, y=ber_DF]{./fig/20220912170545_large_non_sys_uncoded_QPSK/BER_estimators_DetNet.csv};
\addlegendentry{\scriptsize DFE}
\end{semilogyaxis}
\end{tikzpicture}
\vspace{-0.3cm}
\caption{BER performance comparison for system~II, uncoded case.}
\label{fig:system_II_non_sys_uncoded_QPSK}
\end{center}
\vspace{-0.3cm}
\end{figure}

We start the performance comparison of the NN-based equalizers by highlighting the importance of data pre-processing. Without data normalization, the NNs exhibit even worse performance than the LMMSE estimator, which is exemplarily shown for DetNet in Fig.~\ref{fig:system_I_non_sys_uncoded_QPSK} (dotted line). Utilizing the normalized data leads to a major performance improvement (dashed line in Fig.~\ref{fig:system_I_non_sys_uncoded_QPSK}). For DetNet, the BER performance can be further boosted by employing preconditioning, such that with this NN close to optimal MMSE performance can be achieved. The FCNN performs approximately equivalently to the DetNet without preconditioning, while the Attention Detector can outperform the FCNN, which confirms the idea of exploiting correlations for enhancing the estimation performance by utilizing the self-attention mechanism. It turns out, that the SNR utilized for the sample transmission contained in the training set has a large influence on the performance of the NN-based data estimators. Training at too low SNRs leads to flattening out BER curves of the NN-based data estimators at higher SNRs. Training solely at higher SNRs, in turn, impairs the overall performance of the NNs, which probably comes from too few data samples located around the optimal decision boundaries (these samples are very important for the NNs to learn good decision boundaries). Hence, the $E_b/N_0$ training range is another hyperparameter for the NN-based data estimators, whereby the $E_b/N_0$ values for the data burst transmissions contained in the training set are chosen randomly, with uniform distribution on a linear scale within the specified range. For system~I, all NNs are trained in the $E_b/N_0$ range $[9\,\si{dB}, 18\,\si{dB}]$.

Regarding the model-based equalizers, we observe a large performance gap between the LMMSE estimator and other estimators. With the DFE, a performance close to the optimal MMSE performance can be achieved, while the BER performance difference between the vector ML estimator and the MMSE estimator is negligible for the considered system.

As illustrated in Fig.~\ref{fig:system_II_non_sys_uncoded_QPSK}, for system~II, DetNet can slightly outperform the DFE. Similar as for system~I, the Attention Detector exhibits a small performance gap compared to the DetNet, while it outperforms the FCNN. All NNs considered clearly outperform the LMMSE baseline performance.
While for DetNet the $E_b/N_0$ training range is chosen to be $[18\,\si{dB}, 27.5\,\si{dB}]$, the Attention Detector and the FCNN exhibit better performance for an $E_b/N_0$ training range of $[15\,\si{dB}, 27.5\,\si{dB}]$. 

\subsection{Bit Error Ratio Performance -- Coded Transmission}
\label{ssec:Bit_Error_Ratio_Performance_Coded}

\begin{figure*}[t]
\centering
\subfloat[DetNet\label{subfig:Distribution_LLRs_4dB_DetNet}]{
\centering
\includegraphics[width=4.3cm]{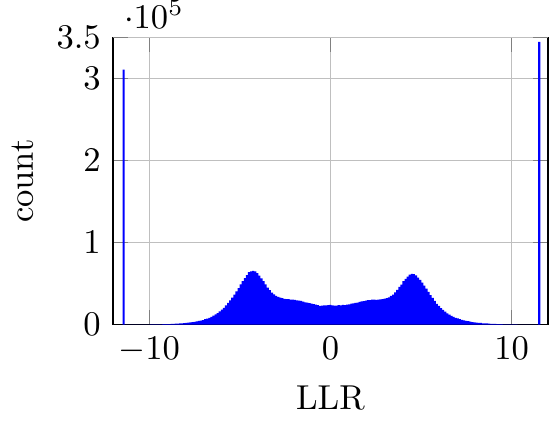}
\vspace{-0.6cm}
}\hspace{0.3cm}
\subfloat[LMMSE\label{subfig:Distribution_LLRs_4dB_LMMSE}]{
\centering
\includegraphics[width=4.3cm]{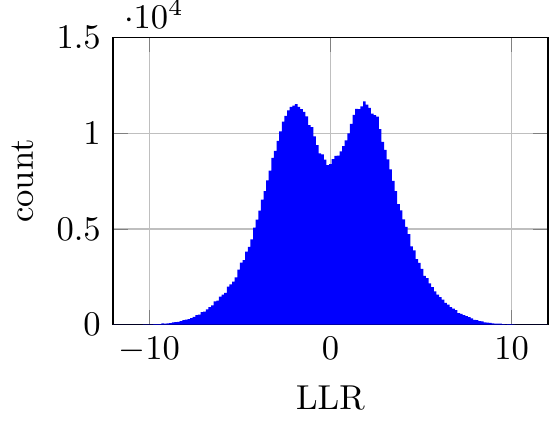}
\vspace{-0.6cm}
}\hspace{0.3cm}
\subfloat[Empirical distrib. true LLRs\label{subfig:Distribution_LLRs_4dB_MMSE}]{
\centering
\includegraphics[width=4.3cm]{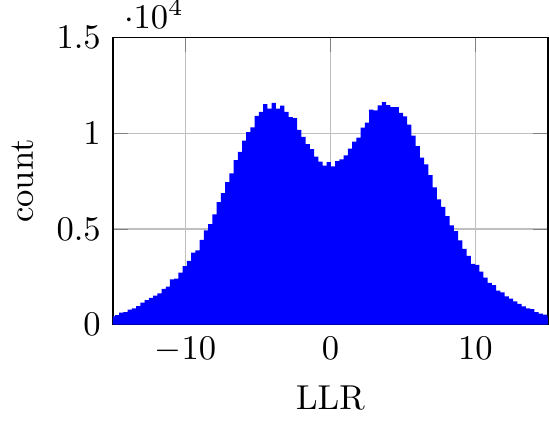}
\vspace{-0.6cm}
}
\caption{Empirical distribution of the LLRs provided by DetNet trained in an $E_b/N_0$ range of $[1\,\si{dB}, 9\,\si{dB}]$~(a) and by the LMMSE estimator~(b), compared with the empirical distribution of the true LLRs~(c) at $E_b/N_0 = 4\,\si{dB}$.}
\label{fig:Distribution_LLRs_4dB}
\end{figure*}

\begin{figure}[!t]
\centering
\subfloat[$E_b/N_0 = 4\,\si{dB}$]{
\centering
\includegraphics[width=4.2cm]{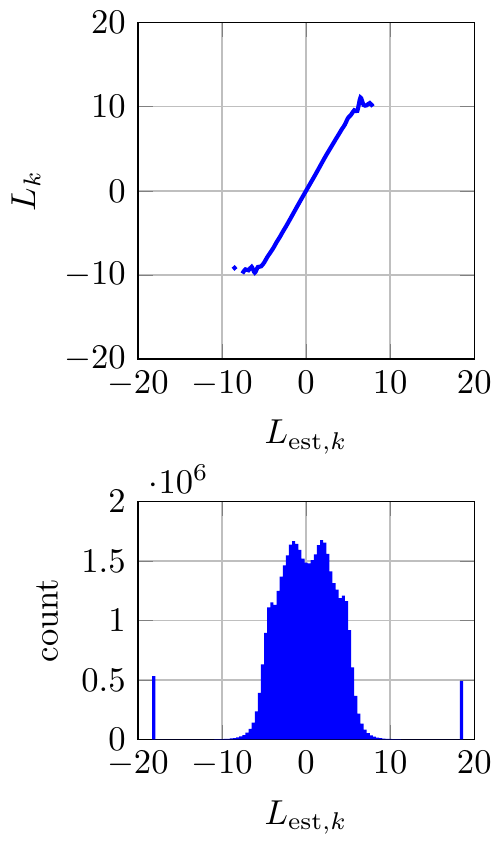}
}
\subfloat[$E_b/N_0 = 8\,\si{dB}$]{
\centering
\includegraphics[width=4.2cm]{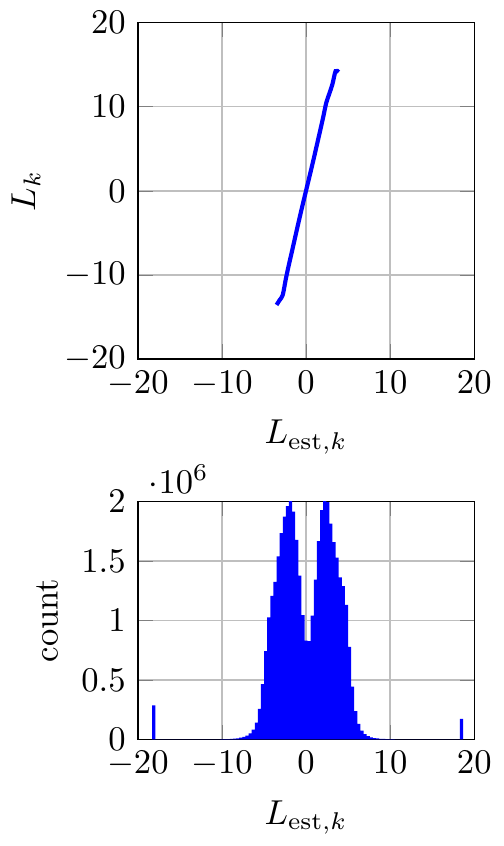}
}
\caption{Distribution of the estimated LLRs provided by the DetNet trained at $1.5\,\si{dB}$, and their relation to the empirical LLRs at a test SNR of (a) $E_b/N_0 = 4\,\si{dB}$ and (b) $E_b/N_0 = 8\,\si{dB}$.}
\label{fig:Distribution_and_LLR_relation}
\end{figure}

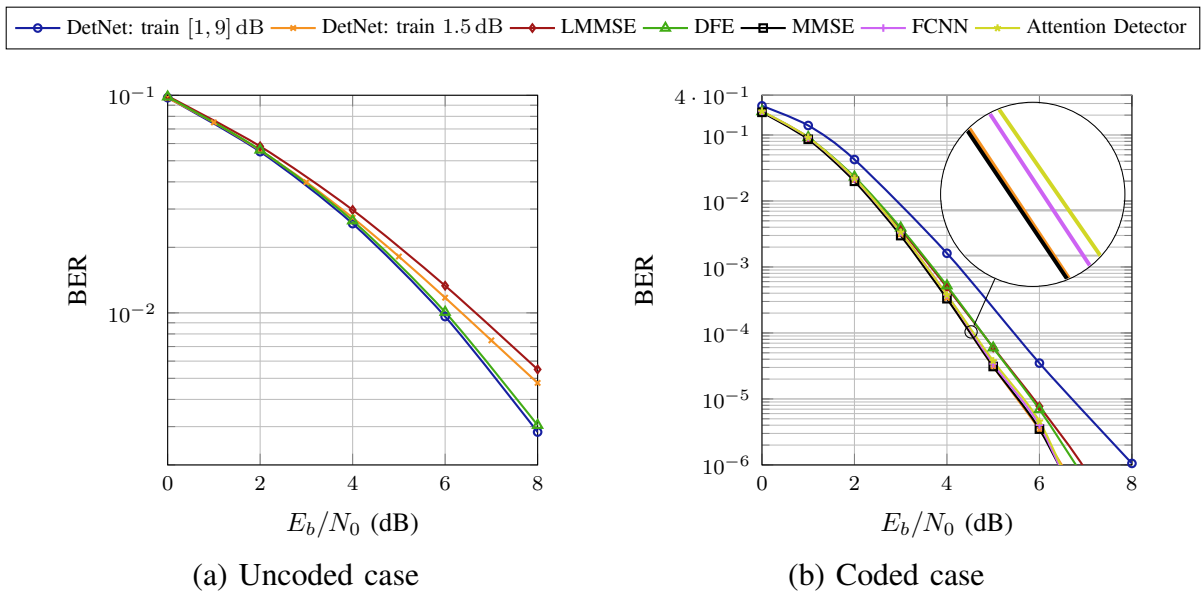
\begin{figure*}[t]
\begin{minipage}{\textwidth}
\centering
\ref{common_legend}
\end{minipage}
\begin{minipage}{\textwidth}
\centering
\subfloat[Uncoded case\label{subfig:system_I_non_sys_uncoded_QPSK_low_SNR}]{
\centering
\begin{tikzpicture}
\begin{semilogyaxis}[compat=newest, width=6.5cm, height=6.5cm, grid=both, ylabel={\small BER}, ymax = 1e-1, ymin = 2e-3, xmax=8, xmin=0, xlabel={\small $E_b/N_0$ (dB)}, scaled ticks = false,  x tick label style={/pgf/number format/.cd,fixed,precision=2,/tikz/.cd}, ticklabel style={font=\scriptsize}, every axis plot/.append style={thick}]
\addplot[color=myblue, mark=o, smooth, mark size=0.05cm] table[col sep=semicolon, x=EbN0_dB, y=ber_DetNet]{fig/20210322221240_small_non_sys_uncoded_QPSK_low_SNRs/BER_estimators.csv};
\addplot[color=myorange, mark=x, smooth, mark size=0.05cm] table[col sep=semicolon, x=EbN0_dB, y=ber_DetNet]{fig/20211013085101_small_non_sys_uncoded_QPSK_low_SNR_better_NN_training/BER_estimators.csv};
\addplot[color=myred, mark=diamond, smooth, mark size=0.05cm] table[col sep=semicolon, x=EbN0_dB, y=ber_lin_est]{fig/20210322221240_small_non_sys_uncoded_QPSK_low_SNRs/BER_estimators.csv};
\addplot[color=mygreen, mark=triangle, smooth] table[col sep=semicolon, x=EbN0_dB, y=ber_DF]{fig/20210322221240_small_non_sys_uncoded_QPSK_low_SNRs/BER_estimators.csv};
\end{semilogyaxis}
\end{tikzpicture}
}\hspace{0.5cm}
\subfloat[Coded case\label{subfig:system_I_non_sys_coded_QPSK}]{
\centering
\begin{tikzpicture}[
    spy/.style={%
        draw,black,
        line width=0.4pt,
        circle,inner sep=0pt,
    },
]

    \def\spyviewersize{2.45cm}

    \def\spyonclipreduce{0.4pt}

    \def\spyfactorI{15}
    \coordinate (spy-in 1) at (3.6,3.6);
    \coordinate (spy-on 1) at (2.78,1.77);

\def\pik{
\begin{semilogyaxis}[compat=newest, width=6.5cm, height=6.5cm, grid=both, ylabel={\small BER}, ymax = 4e-1, ymin = 1e-6, xmax=8, xmin=0, xlabel={\small $E_b/N_0$ (dB)}, scaled ticks = false,  x tick label style={/pgf/number format/.cd,fixed,precision=2,/tikz/.cd}, ticklabel style={font=\scriptsize}, extra y ticks={4e-1}, extra y tick labels={$4\cdot 10^{-1}$}, legend columns = {7}, legend cell align=left, legend style={font=\scriptsize, at={(1, 1.01)}, anchor=south}, legend to name={common_legend}, every axis plot/.append style={thick}]
\addplot[color=myblue, mark=o, smooth, mark size=0.05cm] table[col sep=semicolon, x=EbN0_dB, y=ber_DetNet]{fig/20210317113908_small_non_sys_coded_QPSK/BER_estimators_mod.csv};
\addlegendentry{\scriptsize DetNet: train $[1, 9]\,\si{dB}$}
\addplot[color=myorange, mark=x, smooth, mark size=0.05cm] table[col sep=semicolon, x=EbN0_dB, y=ber_DetNet]{fig/20210915072437_small_non_sys_coded_QPSK_better_NN_training/BER_estimators_DetNet.csv};
\addlegendentry{\scriptsize DetNet: train $1.5\,\si{dB}$}
\addplot[color=myred, mark=diamond, smooth, mark size=0.05cm] table[col sep=semicolon, x=EbN0_dB, y=ber_lin_est]{fig/20210915072437_small_non_sys_coded_QPSK_better_NN_training/BER_estimators_DetNet.csv};
\addlegendentry{\scriptsize LMMSE}
\addplot[color=mygreen, mark=triangle, smooth] table[col sep=semicolon, x=EbN0_dB, y=ber_DF]{fig/20210915072437_small_non_sys_coded_QPSK_better_NN_training/BER_estimators_DetNet.csv};
\addlegendentry{\scriptsize DFE}
\addplot[color=black, mark=square, smooth, mark size=0.05cm] table[col sep=semicolon, x=EbN0_dB, y=ber_MMSE]{fig/20210915072437_small_non_sys_coded_QPSK_better_NN_training/BER_estimators_DetNet.csv};
\addlegendentry{\scriptsize MMSE}
\addplot[color=myviolet, mark=+, smooth, mark size=0.05cm] table[col sep=semicolon, x=EbN0_dB, y=ber_FCNNCompressedInp]{fig/20210915072437_small_non_sys_coded_QPSK_better_NN_training/BER_estimators_AttDet_CompressedInpFCNN.csv};
\addlegendentry{\scriptsize FCNN}
\addplot[color=mykaki, mark=star, smooth, mark size=0.05cm] table[col sep=semicolon, x=EbN0_dB, y=ber_AttentionDetector]{fig/20210915072437_small_non_sys_coded_QPSK_better_NN_training/BER_estimators_AttDet_CompressedInpFCNN.csv};
\addlegendentry{\scriptsize Attention Detector}
\end{semilogyaxis}}
\pik

\def\pik2{
\begin{semilogyaxis}[compat=newest, width=6.5cm, height=6.5cm, grid=both, minor grid style={line width=0.05pt}, major grid style={line width=0.05pt}, ymax = 4e-1, ymin = 1e-6, xmax=8, xmin=0, scaled ticks = false,  x tick label style={/pgf/number format/.cd,fixed,precision=2,/tikz/.cd}, ticklabel style={font=\scriptsize}, extra y ticks={4e-1}, extra y tick labels={$4\cdot 10^{-1}$}, every axis plot/.append style={line width=0.1pt}]
\addplot[color=myblue, mark=o, smooth, mark size=0.05cm] table[col sep=semicolon, x=EbN0_dB, y=ber_DetNet]{fig/20210317113908_small_non_sys_coded_QPSK/BER_estimators_mod.csv};
\addplot[color=myorange, mark=x, smooth, mark size=0.05cm] table[col sep=semicolon, x=EbN0_dB, y=ber_DetNet]{fig/20210915072437_small_non_sys_coded_QPSK_better_NN_training/BER_estimators_DetNet.csv};
\addplot[color=myred, mark=diamond, smooth, mark size=0.05cm] table[col sep=semicolon, x=EbN0_dB, y=ber_lin_est]{fig/20210915072437_small_non_sys_coded_QPSK_better_NN_training/BER_estimators_DetNet.csv};
\addplot[color=mygreen, mark=triangle, smooth] table[col sep=semicolon, x=EbN0_dB, y=ber_DF]{fig/20210915072437_small_non_sys_coded_QPSK_better_NN_training/BER_estimators_DetNet.csv};
\addplot[color=black, mark=square, smooth, mark size=0.05cm] table[col sep=semicolon, x=EbN0_dB, y=ber_MMSE]{fig/20210915072437_small_non_sys_coded_QPSK_better_NN_training/BER_estimators_DetNet.csv};
\addplot[color=myviolet, mark=+, smooth, mark size=0.05cm] table[col sep=semicolon, x=EbN0_dB, y=ber_FCNNCompressedInp]{fig/20210915072437_small_non_sys_coded_QPSK_better_NN_training/BER_estimators_AttDet_CompressedInpFCNN.csv};
\addplot[color=mykaki, mark=star, smooth, mark size=0.05cm] table[col sep=semicolon, x=EbN0_dB, y=ber_AttentionDetector]{fig/20210915072437_small_non_sys_coded_QPSK_better_NN_training/BER_estimators_AttDet_CompressedInpFCNN.csv};
\end{semilogyaxis}}

    \node[spy,minimum size={\spyviewersize/\spyfactorI}] (spy-on node 1) at (spy-on 1) {};
    \node[spy,minimum size=\spyviewersize] (spy-in node 1) at (spy-in 1) {};
    \begin{scope}
        \clip (spy-in 1) circle (0.5*\spyviewersize-\spyonclipreduce);
        \node[circle, fill=white, draw=white, minimum size=\spyviewersize] (background_circ) at (spy-in 1) {};
        \pgfmathsetmacro\sI{1/\spyfactorI}
        \begin{scope}[
            shift={($\sI*(spy-in 1)-\sI*(spy-on 1)$)},
            scale around={\spyfactorI:(spy-on 1)}
        ]
            \pik2
        \end{scope}
    \end{scope}
    \draw [spy] (spy-on node 1) -- (spy-in node 1);

\end{tikzpicture}
}
\end{minipage}
\caption{Comparison uncoded and coded BER performance for system~I, non-systematic UW-OFDM. DetNet is once trained in an $E_b/N_0$ range of $[1\,\si{dB}, 9\,\si{dB}]$, and once at $E_b/N_0 = 1.5\,\si{dB}$. The FCNN and the Attention Detector are trained at $E_b/N_0 = 0.8\,\si{dB}$.}
\label{fig:system_I_non_sys_coded_uncoded}
\end{figure*}

As already described in Sec.~\ref{sec:Neural_Network_Based_Data_Estimation}, the NNs are trained to provide estimates for the posterior probabilities for every data symbol estimate for both coded and uncoded data transmission. That is, we expect a trained NN-based data estimator to be applicable for coded and uncoded transmission without requiring retraining. As detailed in Sec.~\ref{ssec:Bit_Error_Ratio_Performance_Uncoded}, for uncoded transmission it is beneficial to train the NNs for different SNRs, where the SNR training range limits can be viewed as hyperparameters -- with this approach a good, or even close to optimal BER performance can be achieved. However, employing these trained NNs for coded transmission, their performance is unsatisfactory. As shown in Fig.~\ref{subfig:system_I_non_sys_coded_QPSK} exemplarily for DetNet, the NN-based equalizer trained in an $E_b/N_0$ range of $[1\,\si{dB}, 9\,\si{dB}]$ performs distinctly worse than the DFE and the LMMSE estimator, while the same NN outperforms both model-based equalizers for uncoded transmission (Fig.~\ref{subfig:system_I_non_sys_uncoded_QPSK_low_SNR}). The reason for this result can be explained by investigating the empirical distribution of the LLRs provided by DetNet. Comparing the  LLRs of DetNet trained in an $E_b/N_0$ range of $[1\,\si{dB}, 9\,\si{dB}]$ (Fig.~\ref{subfig:Distribution_LLRs_4dB_DetNet}) with the true LLRs at $E_b/N_0 = 4\,\si{dB}$ (Fig.~\ref{subfig:Distribution_LLRs_4dB_MMSE}) reveals that a vast number of LLRs provided by DetNet has a high absolute value\footnote{We introduced an upper and a lower limit for the output values of DetNet since, due to imperfect training, its output values can be slightly smaller than $0$ or greater than $1$, which leads to problems for the computation of the LLRs.}, while this is not the case for the true LLRs and also not for the LLRs of the LMMSE estimator (Fig.~\ref{subfig:Distribution_LLRs_4dB_LMMSE}). That is, the NN is overconfident in many of its decisions, which harms the performance of the Viterbi channel decoder. 

To tackle this problem, we investigated treating the data estimation problem as a classification task, i.e., we utilized Softmax as an output activation function of the NNs, combined with using cross-entropy loss for training. Then, so-called label smoothing can be applied, which is a common approach for combating overconfidence of classification NNs~\cite{Mueller19}. Unfortunately, this approach did not lead to significant performance improvements in our experiments. However, we observed that the training $E_b/N_0$ range has a large impact on the distribution of the LLRs provided by DetNet. More specifically, the overconfidence of an NN-based equalizer can be highly reduced by training at low SNRs. This highlights the importance of the training SNR as a hyperparameter, which has to be chosen differently for coded and uncoded data transmission.

Investigating solely the distribution of the LLRs, however, is only an indicator of their reliability. We utilize an approach described in~\cite{Bauch01} for an assessment of the LLR quality of turbo equalizers. To this end, we apply the trained NNs on the validation set, to obtain the estimated LLRs $L_{\text{est},i}$ for all bits $b_i$ contained in the validation set. The estimated LLRs $L_{\text{est},i}$ are grouped according to their value into $K$ bins with the value $L_k$, $k\in\{0, ..., K-1\}$ ($L_k$ is the mean of the estimated LLRs in bin $k$). The signs of $L_{\text{est},i}$ are used for a hard decision estimate of the corresponding bits $b_i$. With these hard decision estimates at hand, the empirical bit error probability 
\begin{gather*}
P_{\text{emp},k} = \frac{\#\,\text{wrong hard decisions in bin}\,k}{\#\,\text{bits in bin}\,k}
\end{gather*} can be computed for all $K$ bins. These empirical bit error probabilities, in turn, can be utilized to determine the empirical LLRs $L_{\text{emp},k}$ for all $K$ bins with
\begin{gather}
L_{\text{emp},k} = \text{sign}(L_k)\left|\ln\left(\frac{1-P_{\text{emp},k}}{P_{\text{emp},k}}\right)\right|\,.
\end{gather}
Assuming a sufficiently large number of LLR values per bin, the empirical LLRs $L_{\text{emp},k}$ provide an approximation of the true LLRs. The quality of the estimated LLRs $L_{\text{est},i}$ can be ascertained by plotting $L_{\text{emp},k}$ against $L_k$. Since the estimated LLRs should match the empirical ones, the plotted graph is ideally a linear function with slope one. However, also slopes not equal to one allow optimal channel decoding performance of the Viterbi decoder, since all LLRs are under- or overrated in the same fashion. Nonlinear graphs, in turn, indicate a loss in BER performance, since some estimated LLRs are overrated while others are underrated at the same time. This may lead to wrong decisions of the Viterbi channel decoder when searching the optimum path in the trellis diagram of the convolutional code. As shown in Fig.~\ref{fig:Distribution_and_LLR_relation} for the LLRs provided by DetNet when being trained at $1.5\,\si{dB}$, the number of LLRs with too high value could be drastically lowered. Further, the empirical LLRs and the estimated LLRs are related nearly linearly for the majority of the estimated LLRs, i.e., in the regions where the relation is nonlinear, the counts per LLR bin are comparatively small.

As the BER curves in Fig.~\ref{subfig:system_I_non_sys_coded_QPSK} show, DetNet trained at $1.5\,\si{dB}$ achieves close to optimal BER performance. However, for uncoded transmission, the DetNet trained at $E_b/N_0 = 1.5\,\si{dB}$ performs distinctly worse than the DetNet trained in the $E_b/N_0$ range of $[1\,\si{dB}, 9\,\si{dB}]$, which is depicted in Fig.~\ref{subfig:system_I_non_sys_uncoded_QPSK_low_SNR}. For the Attention Detector and the FCNN, $E_b/N_0 = 0.8\,\si{dB}$ is utilized as an SNR for training, all other hyperparameters are chosen as for uncoded data transmission. Both achieve close to optimal BER performance, too. 

\begin{figure}[t]
\begin{center}
\begin{tikzpicture}
\begin{semilogyaxis}[compat=newest, width=6.5cm, height=6.5cm, grid=both, ylabel={\small BER}, ymax = 5e-1, ymin = 1e-6, xmax=12, xmin=0, xlabel={\small $E_b/N_0$ (dB)}, scaled ticks = false,  x tick label style={/pgf/number format/.cd,fixed,precision=2,/tikz/.cd}, ticklabel style={font=\scriptsize}, extra y ticks={5e-1}, extra y tick labels={$5\cdot 10^{-1}$}, xtick={0, 2, 4, 6, 8, 10, 12}, legend columns = {1}, legend cell align=left, legend style={font=\scriptsize}, every axis plot/.append style={thick}]
\addplot[color=myblue, mark=o, smooth] table[col sep=semicolon, x=EbN0_dB, y=ber_DetNet]{./fig/20211103130924_large_non_sys_coded_QPSK/BER_estimators.csv};
\addlegendentry{\scriptsize DetNet}
\addplot[color=myred, mark=diamond, smooth] table[col sep=semicolon, x=EbN0_dB, y=ber_lin_est]{./fig/20211103130924_large_non_sys_coded_QPSK/BER_estimators.csv};
\addlegendentry{\scriptsize LMMSE}
\addplot[color=mygreen, mark=triangle, smooth] table[col sep=semicolon, x=EbN0_dB, y=ber_DF]{./fig/20211103130924_large_non_sys_coded_QPSK/BER_estimators.csv};
\addlegendentry{\scriptsize DFE}
\end{semilogyaxis}
\end{tikzpicture}
\vspace{-0.3cm}
\caption{BER performance comparison for system II, coded case.}
\label{fig:system_II_non_sys_coded_QPSK}
\end{center}
\end{figure}
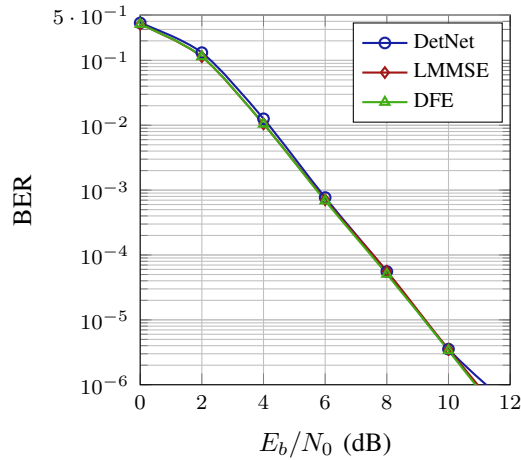

For system~II, we compare the LMMSE estimator, the DFE, and the DetNet, which is trained at $E_b/N_0 = 4\,\si{dB}$. As shown in Fig.~\ref{fig:system_II_non_sys_coded_QPSK}, all three investigated equalizers exhibit approximately the same BER performance for coded data transmission. Although simulating the optimal BER performance is computationally infeasible, it can be stated that the achieved performance of the three equalizers is very close to the optimal performance. This statement can be verified by considering the LLRs provided by the LMMSE estimator. They are equivalent to the true LLRs when the conditional distribution $p(\hat{d}_i^\prime|d_i^\prime)$ is Gaussian (cf. Sec.~\ref{ssec:LMMSE}). Since this condition is well fulfilled for the system dimensions of system~II, the LLRs of the LMMSE are close to the true LLRs, leading to close to optimal BER performance for coded data transmission.

\subsection{Distributions of the Data Estimates}
\label{ssec:Distributions_of_the_Data_Estimates}
\begin{figure*}[t]
\centering
\subfloat[LMMSE\label{subfig:Distribution_cond_data_estim_4dB_LMMSE}]{
\centering
\includegraphics[width=0.23\textwidth]{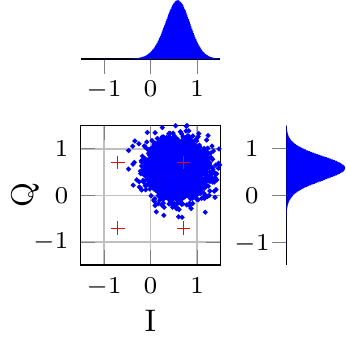}
\vspace{-0.3cm}
}\hspace{0.3cm}
\subfloat[MMSE\label{subfig:Distribution_cond_data_estim_4dB_MMSE}]{
\centering
\includegraphics[width=0.23\textwidth]{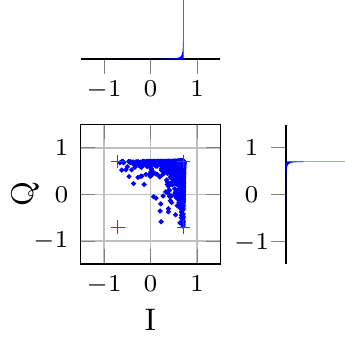}
\vspace{-0.3cm}
}\hspace{0.3cm}
\subfloat[DetNet\label{subfig:Distribution_cond_data_estim_4dB_DetNet}]{
\centering
\includegraphics[width=0.23\textwidth]{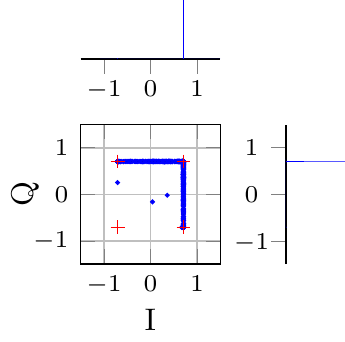}
\vspace{-0.3cm}
}
\caption{Distribution of the conditional data symbol estimates for system~I at $E_b/N_0 = 4\,\si{dB}$.}
\label{fig:Distribution_cond_data_estim_4dB}
\end{figure*}

We also want to highlight the differences in the distributions of the estimates of the MMSE estimator, the NN-based estimators (exemplarily shown for DetNet), and the LMMSE estimator. To this end, we visualize the conditional distributions of their estimates, given a transmitted symbol $(1+j)/\sqrt{2} $, for system~I at $E_b/N_0 = 4\,\si{dB}$ in I/Q diagrams. The empirical distributions of the data symbol estimates are plotted in histograms along the I-axis and the Q-axis. As shown in Fig.~\ref{subfig:Distribution_cond_data_estim_4dB_LMMSE}, the conditional LMMSE estimates follow, as expected, (approximately) a Gaussian distribution. However, the MMSE estimates are distributed in a completely different manner. As indicated by the histograms in Fig.~\ref{subfig:Distribution_cond_data_estim_4dB_MMSE}, the vast majority of the estimates are located very close to the constellation point. Since the MMSE estimator yields the posterior expectation of a data symbol as an estimate, no estimate can lie outside the square connecting the four constellation points (marked by red crosses). The estimates of DetNet, plotted in Fig.~\ref{subfig:Distribution_cond_data_estim_4dB_DetNet}, exhibit a distribution similar to that of the MMSE estimates. This is in fact expected, since, due to training the NNs with a quadratic loss function, the NNs try to minimize the cost metric that the MMSE estimator minimizes, namely the Bayesian mean square error. Hence, the trained NNs approximate the MMSE estimator function.

\subsection{Complexity Analysis}
\label{ssec:Complexity_Analysis}
In this section, we provide a brief analysis of the inference complexity of the presented NN-based data estimators as well as of the LMMSE estimator and the DFE in terms of the number of required scalar, real-valued multiplications needed for equalization of one UW-OFDM data symbol. In this paper, we account four real-valued multiplications for one complex-valued multiplication. Data normalization, as well as the complexity required for training the NNs is not regarded in this analysis.

For DetNet, we first determine the complexity of a single layer. Given $\m{H}^T\m{H}$, the number of multiplications carried out in a layer according to~\eqref{eq:DetNet_linear_step} including the projection by an FCNN with a single hidden layer, one-hot demapping, and the weighted residual connections is
\begin{gather}
\begin{aligned}
M_{\text{DetNet},k} &= \underbrace{4N_{\text{d}}^2}_{\m{H}^T\m{H}\hat{\ve{d}}_k} + \underbrace{2d_{\text{h}}(N_{\text{d}}(|\mathbb{S}|+1)+d_{\text{v}})}_{\text{single hid. layer FCNN}}+ \underbrace{2N_{\text{d}}}_{\delta_{1k}\cdot} + \underbrace{2N_{\text{d}}}_{\delta_{2k}\cdot} +\underbrace{2N_{\text{d}}|\mathbb{S}|}_{\text{one-hot demap.}}+\underbrace{2N_{\text{d}}+d_{\text{v}}}_{\text{residual}}\,.
\end{aligned}
\end{gather}
Overall, DetNet has an inference complexity of 
\begin{gather}
\begin{aligned}
M_{\text{DetNet}} &= L M_{\text{DetNet},k} - 2N_{\text{d}}|\mathbb{S}| + 8N_{\text{d}}^2(N_{\text{d}}+N_{\text{u}})+4N_{\text{d}}(N_{\text{d}}+N_{\text{u}}) + 2N_{\text{d}}(2N_{\text{d}}+1)
\end{aligned}
\end{gather}
real-valued multiplications, where we consider with the subtracted term that no one-hot decoding is conducted in the last layer, while with the three added terms the computations of $\m{H}^T\m{H}$ and of $\m{H}^T\ve{y}$, and the preconditioning are taken into account.

Determining the number of multiplications required by the FCNN is straightforward and can be expressed as
\begin{gather}
\begin{aligned}
M_{\text{FCNN}} &= \underbrace{(2N_{\text{d}}^2 + 2N_{\text{d}})d_{\text{h}}}_{\text{input to hidden}}+\underbrace{d_{\text{h}}^2(L-1)}_{\text{hidden to hidden}} + \underbrace{2d_{\text{h}}N_{\text{d}}|\mathbb{S}|}_{\text{hidden to output}} + 2(2N_{\text{d}}^2 + N_{\text{d}})(N_{\text{d}}+N_{\text{u}}) + 4N_{\text{d}}(N_{\text{d}}+N_{\text{u}})\,,
\end{aligned}
\end{gather}
where with the last two terms the computations of the upper triangular matrix (including the main diagonal) of $\m{H}^T\m{H}$ and of $\m{H}^T\ve{y}$ are considered. 

For the Attention Detector, we start by evaluating the complexity of a single encoder layer, which consists of a self-attention layer, a single hidden layer FCNN, layer normalization, and residual connections. The inputs of the self-attention layer are mapped to so-called queries~$\ve{q}_i$, keys~$\ve{k}_i$, and values~$\ve{v}_i$, $i\in\{0, ..., 2N_{\text{d}}-1\}$, by multiplying with learned matrices~\cite{Vaswani17}. Then, self-attention scores between each query and each key are computed, followed by a weighting of the values~$\ve{v}_i$ by these scores. The number of multiplications conducted in one encoder layer thus is 
\begin{gather}
\begin{aligned}
M_{\text{AttEnc},k} &= \underbrace{6N_{\text{d}}(2N_{\text{d}}+1)^2}_{\text{input mappings}} + \underbrace{4N_{\text{d}}^2(4N_{\text{d}} + 3)}_{\text{score weighting}} +\underbrace{8N_{\text{d}}(2N_{\text{d}}+1)}_{\text{normalization, residual}}+\underbrace{4N_{\text{d}}d_{\text{h,enc}}(2N_{\text{d}}+1)}_{\text{single hid. layer FCNN}}\,.
\end{aligned}
\end{gather}
For the FCNN on top of the encoder, another
\begin{gather}
\begin{aligned}
M_{\text{AttFcnn}} &= \underbrace{2N_{\text{d}}d_{\text{h,fcnn}}(2N_{\text{d}}+1)}_{\text{input to hidden}} +\underbrace{(L_{\text{fcnn}}-1)d_{\text{h,fcnn}}^2}_{\text{hidden to hidden}} + \underbrace{2N_{\text{d}}|\mathbb{S}|d_{\text{h,fcnn}}}_{\text{hidden to output}}
\end{aligned}
\end{gather}
multiplications have to be performed. Therefore, the complexity of the Attention Detector is given by
\begin{gather}
\begin{aligned}
M_{\text{AttDet}} &= L_{\text{enc}}M_{\text{AttEnc},k}+M_{\text{Att,fcnn}} + 8N_{\text{d}}^2(N_{\text{d}}+N_{\text{u}})+4N_{\text{d}}(N_{\text{d}}+N_{\text{u}}) + 2N_{\text{d}}(2N_{\text{d}}+1)\,,
\end{aligned}
\end{gather}
where we consider the computations of $\m{H}^T\m{H}$ and $\m{H}^T\ve{y}$, as well as the preconditioning  with the last three terms. 

For the LMMSE estimator, we first regard the complexity for obtaining the estimator matrix $\m{E}_{\text{LMMSE}}$. Since the channel is assumed to be stationary for a whole data burst, the estimator matrix has to be computed only once per burst. Assuming that the inversion in~\eqref{eq:LMMSE_estimator} is computed by a Cholesky decomposition as of~\cite{Golub13}, the computation of $\m{E}_{\text{LMMSE}}$ entails a complexity of 
\begin{align}
M_{\text{LMMSE,burst}}&=\underbrace{\frac{14}{3}N_{\text{d}}^3+ 4N_{\text{d}}^2}_{\text{inverse (Cholesky)}}  + \underbrace{8N_{\text{d}}^2(N_{\text{d}} + N_{\text{u}})}_{\text{multiplication with $\m{H}^{\prime H}$}} =\frac{38}{3}N_{\text{d}}^3 + 8N_{\text{d}}^2N_{\text{u}} + 4N_{\text{d}}^2\,.
\end{align}
Then, given $\m{E}_{\text{LMMSE}}$, the number of required multiplications for the equalization of every received UW-OFDM vector is
\begin{gather}
M_{\text{LMMSE,eq}} = 4(N_{\text{d}} + N_{\text{u}})N_{\text{d}}\,.
\end{gather}

We determine the DFE complexity in Alg.~\ref{alg:DFE} by first considering those computations that have to be done once for every data burst. Namely, this refers to the computations of the estimator vectors $\ve{e}_k^H$ and the error covariance matrices $\m{C}_{\text{ee},k}$ for every iteration step. We note that $\m{H}^{\prime H}\m{H}$ needs to be computed only once, and then the matrices $\m{H}_k^{\prime H}\m{H}_k$ can be retrieved by deleting the appropriate rows and columns. The size of $\m{H}_k^\prime$ decrements in every iteration, and thus we elaborate the complexity of computing $\m{A}_k$ given $\m{H}_k^{\prime H}\m{H}_k\in\mathbb{C}^{C\times C}$, with $C\in\{2, ..., N_{\text{d}}\}$. Furthermore, the scaling of $\m{A}_k$ by $N\sigma_{\text{n}}^2$, as described in Alg.~\ref{alg:DFE}, line~\ref{alg:DFE_Cee}, can be omitted, since only the minimum value on the diagonal of $\m{C}_{\text{ee},k}$ is needed for finding the index $j$. In summary,
\begin{align}
M_{\text{DFE,burst}} &= \underbrace{4N_{\text{d}}^2(N_{\text{d}} + N_{\text{u}})}_{\m{H}^{\prime H}\m{H}^{\prime}} + \underbrace{4(N_{\text{d}} + N_{\text{u}} + 1)}_{\text{last estimator vector}} + \sum_{C=2}^{N_{\text{d}}}\underbrace{\frac{14}{3}C^3 + 4C^2}_{\m{A}_k \text{\,(Cholesky)}} + \underbrace{4C(N_{\text{d}} + N_{\text{u}})}_{\ve{e}_k^H}\nonumber\\
&= \frac{7}{6}N_{\text{d}}^4 + \frac{29}{3}N_{\text{d}}^3 + \frac{31}{6}N_{\text{d}}^2 + 6N_{\text{d}}^2N_{\text{u}}+ \frac{2}{3}N_{\text{d}} + 2N_{\text{d}}N_{\text{u}} - \frac{14}{3}
\end{align}
multiplications have to be carried out once for every data burst to obtain the $N_{\text{d}}$ estimator vectors $\ve{e}_k^H$. 
For both the estimation of a single data symbol and the removal of the influence of this estimate on the received vector, $(N_{\text{d}} + N_{\text{u}})$ complex-valued multiplications have to be accounted for. Hence, given the estimator vectors, equalization of every received UW-OFDM vector with the DFE has a complexity of
\begin{gather}
M_{\text{DFE,eq}} = 8N_{\text{d}}^2 + 8N_{\text{d}}N_{\text{u}}\,.
\end{gather} 

The particular complexity numbers of the considered equalizers are stated in Tab.~\ref{tab:Nr_mult_equalizers} for both system~I and system~II. Obviously, the NN-based equalizers exhibit a distinctly higher complexity than the considered model-based ones. However, a comparison of the complexities of the DetNet and the DFE reveals that the complexity of the DFE grows significantly faster with the dimension of the UW-OFDM system model than that of the DetNet. Among the considered NNs, the DetNet is the lowest complex equalizer. That is, incorporating model knowledge directly into the layers structure of an NN seems to be most promising for obtaining well-performing and comparably low complex NN-based data estimators.

{
\renewcommand{\aboverulesep}{0pt}
\renewcommand{\belowrulesep}{0pt}
\renewcommand{\arraystretch}{1.3}
\begin{table}[t]
\caption{Number of required multiplications of considered equalizers rounded to hundreds.}
\label{tab:Nr_mult_equalizers}
\begin{center}
\begin{tabularx}{12cm}{>{\raggedright}m{2.5cm} >{\raggedright}m{1.8cm} >{\raggedleft}X >{\raggedleft\arraybackslash}X}
\toprule
& & \centering System I & \centering System II\tabularnewline
\hline
DetNet & $M_{\text{DetNet}}$ & 100000  & 3178300\\
\hline
FCNN & $M_{\text{FCNN}}$ &  866400& 15437800\\
\hline
Attention Detector & $M_{\text{AttDet}}$ &  614400 & 49971700\\
\hline
\multirow{2}{*}{LMMSE} & $M_{\text{LMMSE,burst}}$ & 8800& 550200\\
\cline{2-4}
& $M_{\text{LMMSE,eq}}$ & 400& 6100\\
\hline
\multirow{2}{*}{DFE} & $M_{\text{DFE,burst}}$ & 11700 & 1644700\\
\cline{2-4}
& $M_{\text{DFE,eq}}$ & 800 & 12300\\
\bottomrule
\end{tabularx}
\end{center}
\vspace{-0.5cm}
\end{table}
}

\section{Conclusion}
\label{sec:Conclusion}
In this paper, we investigated three NN-based approaches for data estimation in UW-OFDM systems, whereby model knowledge was utilized in different ways. Moreover, we described state-of-the-art model-based equalizers, and we discussed the equivalence of the MMSE estimator and the bit-wise MAP estimator for the considered system setup. We pointed out the importance of proper data normalization for NN-based data estimators and proposed a data normalization scheme specifically for UW-OFDM signaling. With preconditioning, we introduced adaptions for DetNet to boost its BER performance and decrease its computational complexity. Further, we showed a model-inspired data pre-processing approach, and we proposed an NN-based data estimator inspired by the Transformer network. We highlighted the difficulties when employing NNs for data estimation in channel coded data transmission, and we introduced a measure for obtaining reliable LLRs by NN-based equalizers. Finally, we provided BER performance results, we conducted a complexity analysis, and we visualized the distribution of the estimates of selected model-based and NN-based equalizers.

{\appendices
\section{Equivalence of the MMSE and the Bit-Wise MAP Hard Decision Estimates for QPSK}
\label{sec:Equivalence_MMSE_Bit-Wise_MAP_Hard_Decision_Estimates}
When employing a QPSK modulation alphabet, the data symbols $d_i^\prime$ are drawn from the modulation alphabet $\mathbb{S}^\prime~:=~\rho\{1+j, 1-j, -1+j, -1-j\}$, where $\rho = 1/\sqrt{2}$ or $\rho = 1$ for a normalized or an unnormalized symbol alphabet, respectively. For deriving the hard decision estimate of the MMSE estimator, we consider the MMSE estimate of the $i$th data symbol, $i\in\{0, ..., N_{\text{d}}-1\}$:
\begin{align}
\hat{d}_i^\prime &= \sum_{\ve{d}^{\prime\prime}\in \mathbb{S}^{\prime N_{\text{d}}}}d_i^{\prime\prime} p[\ve{d}^{\prime\prime}|\ve{y}^\prime]\nonumber\\ 
&=\sum_{d_{i,\text{Re}}^{\prime\prime}\in \mathbb{S}_{\text{Re}}}\sum_{d_{i,\text{Im}}^{\prime\prime}\in \mathbb{S}_{\text{Im}}}(d_{i,\text{Re}}^{\prime\prime}+j d_{i,\text{Im}}^{\prime\prime}) p[(d_{i,\text{Re}}^{\prime\prime}+j d_{i,\text{Im}}^{\prime\prime})|\ve{y}^\prime]\nonumber\\
&= \sum_{d_{i,\text{Re}}^{\prime\prime}\in \mathbb{S}_{\text{Re}}}\sum_{d_{i,\text{Im}}^{\prime\prime}\in \mathbb{S}_{\text{Im}}} d_{i,\text{Re}}^{\prime\prime} p[(d_{i,\text{Re}}^{\prime\prime}+j d_{i,\text{Im}}^{\prime\prime})|\ve{y}^\prime] + j\sum_{d_{i,\text{Re}}^{\prime\prime}\in \mathbb{S}_{\text{Re}}}\sum_{d_{i,\text{Im}}^{\prime\prime}\in \mathbb{S}_{\text{Im}}} d_{i,\text{Im}}^{\prime\prime} p[(d_{i,\text{Re}}^{\prime\prime}+j d_{i,\text{Im}}^{\prime\prime})|\ve{y}^\prime]\nonumber\\
&= \sum_{d_{i,\text{Re}}^{\prime\prime}\in \mathbb{S}_{\text{Re}}}d_{i,\text{Re}}^{\prime\prime} p[d_{i,\text{Re}}^{\prime\prime}|\ve{y}^\prime] + j\sum_{d_{i,\text{Im}}^{\prime\prime}\in \mathbb{S}_{\text{Im}}}d_{i,\text{Im}}^{\prime\prime} p[j d_{i,\text{Im}}^{\prime\prime}|\ve{y}^\prime]\,,\label{eq:MMSE_decision_re_im_independent}
\end{align}
where $\mathbb{S}_{\text{Re}} = \mathbb{S}_{\text{Im}} = \{-\rho,\rho\}$, $d_{i,\text{Re}}^{\prime\prime} := \text{Re}\{d_i^{\prime\prime}\}$, and $d_{i,\text{Im}}^{\prime\prime} := \text{Im}\{d_i^{\prime\prime}\}$. That is, the real and the imaginary part of $d_i$ are estimated independently of each other. Inserting the symbols of the symbol alphabet into~\eqref{eq:MMSE_decision_re_im_independent} leads to
\begin{gather}
\begin{aligned}
\hat{d}_i^\prime &= -\rho p[\text{Re}\{d_i^\prime\} = -\rho|\ve{y}^\prime] + \rho p[\text{Re}\{d_i^\prime\} = \rho|\ve{y}^\prime] + j(- \rho p[\text{Im}\{d_i^\prime\} = -\rho|\ve{y}^\prime] + \rho p[\text{Im}\{d_i^\prime\} = \rho|\ve{y}^\prime])\\
&= -\rho p[b_{0i} = 0|\ve{y}^\prime] + \rho p[b_{0i} = 1|\ve{y}^\prime] + j(- \rho p[b_{1i} = 0|\ve{y}^\prime] + \rho p[b_{1i} = 1|\ve{y}^\prime])\,,
\end{aligned}\label{eq:ith_MMSE_estimate_QPSK}
\end{gather}
where in the last step the QPSK bit-to-symbol mapping described in Sec.~\ref{ssec:Minimum_Mean_Square_Error_Estimator} is applied. 
In case of hard decision,  $\hat{d}_i^\prime$ is sliced to the closest constellation symbol, i.e., $\text{Re}\{\hat{d}_i\}$ is sliced to $\rho$ for $\text{Re}\{\hat{d}_i\} > 0$, and to $-\rho$ otherwise (accordingly for $\text{Im}\{\hat{d}_i\}$). Hence, the real and the imaginary part of an MMSE hard decision estimate $\big\lfloor\hat{d}_i^\prime\big\rceil$ follow to
\begin{gather}
\text{Re}\big\{\big\lfloor\hat{d}_i^\prime\big\rceil\big\} = \begin{cases}
\rho & p[b_{0i} = 1|\ve{y}^\prime] > p[b_{0i} = 0|\ve{y}^\prime] \\
-\rho & \text{otherwise}
\end{cases}\label{eq:MMSE_hard_decision_criterion_re}
\end{gather}
and
\begin{gather}
\text{Im}\big\{\big\lfloor\hat{d}_i^\prime\big\rceil\big\} = \begin{cases}
\rho & p[b_{1i} = 1|\ve{y}^\prime] > p[b_{1i} = 0|\ve{y}^\prime] \\
-\rho & \text{otherwise}
\end{cases},
\label{eq:MMSE_hard_decision_criterion_im}
\end{gather}
respectively, which coincides with a hard decision estimate of the bit-wise MAP estimator. 
}

\bibliography{bibliography.bib}

\end{document}